\documentclass[preprint,pre,aps,showpacs,showkeys,amsmath]{revtex4}
\usepackage{graphicx}

\begin{document}

\title{Fast Sparsely Synchronized Brain Rhythms in A Scale-Free Neural Network}

\author{Sang-Yoon Kim}
\email{sangyoonkim@dnue.ac.kr}
\author{Woochang Lim}
\email{woochanglim@dnue.ac.kr}
\affiliation{Computational Neuroscience Lab., Department of Science Education, Daegu National University of Education, Daegu 705-115, Korea}

\begin{abstract}
We consider a directed version of the Barab\'{a}si-Albert scale-free network (SFN) model with symmetric preferential attachment with the same in- and out-degrees, and study emergence of sparsely synchronized rhythms for a fixed attachment degree in an inhibitory population of fast spiking Izhikevich interneurons. Fast sparsely synchronized rhythms with stochastic and intermittent neuronal discharges are found to appear for large values of $J$ (synaptic inhibition strength) and $D$ (noise intensity). For an intensive study we fix $J$ at a sufficiently large value, and investigate the population states by increasing $D$. For small $D$, full synchronization with the same population-rhythm frequency $f_p$ and mean firing rate (MFR) $f_i$ of individual neurons occurs, while for large $D$ partial synchronization with $f_p > {\langle f_i \rangle}$ ($\langle f_i \rangle$: ensemble-averaged MFR) appears due to intermittent discharge of individual neurons; particularly, the case of $f_p > 4 {\langle f_i \rangle}$ is referred to as sparse synchronization. For the case of partial and sparse synchronization, MFRs of individual neurons vary depending on their degrees. As $D$ passes a critical value $D^*$ (which is determined by employing an order parameter), a transition to unsynchronization occurs due to destructive role of noise to spoil the pacing between sparse spikes. For $D<D^*$, population synchronization emerges in the whole population because the spatial correlation length between the neuronal pairs covers the whole system. Furthermore, the degree of population synchronization is also measured in terms of two types of realistic statistical-mechanical measures. Only for the partial and sparse synchronization, contributions of individual neuronal dynamics to population synchronization change depending on their degrees, unlike the case of full synchronization. Consequently, dynamics of individual neurons reveal the inhomogeneous network structure for the case of partial and sparse synchronization, which is in contrast to the case of statistically homogeneous random graphs and small-world networks. Finally, we investigate the effect of network architecture on sparse synchronization for fixed values of $J$ and $D$ in the following three cases: (1) variation in the degree of symmetric attachment (2) asymmetric preferential attachment of new nodes with different in- and out-degrees (3) preferential attachment between pre-existing nodes (without addition of new nodes). In these three cases, both relation between network topology (e.g., average path length and betweenness centralization) and sparse synchronization and contributions of individual dynamics to the sparse synchronization are discussed.
\end{abstract}


\maketitle

\section{Introduction}
\label{sec:INT}
Recently, brain rhythms in health and disease have attracted much attention \cite{Buz1,TW}. Particularly, we are concerned about fast sparsely synchronized brain rhythms which are related to diverse cognitive functions (e.g., sensory perception, feature integration, selective attention, and memory formation) \cite{W_Review}. At the population level, synchronous small-amplitude fast oscillations (e.g., gamma rhythm (30-100 Hz) during awake behaving states and rapid eye movement sleep and sharp-wave ripple (100-200 Hz) during quiet sleep and awake immobility) have been observed in local field potential recordings, while at the cellular level individual neuronal recordings have been found to exhibit stochastic and intermittent spike discharges like Geiger counters \cite{SS1,SS2,SS3,SS4,SS5,SS6,SS7}. Thus, single-cell firing activity differs distinctly from the population oscillatory behavior.
We note that these sparsely synchronized rhythms are in contrast to fully synchronized rhythms where individual neurons fire regularly at the population frequency like the clocks. Brunel et al. developed a framework appropriate for description of fast sparse synchronization \cite{Sparse1,Sparse2,Sparse3,Sparse4,Sparse5,Sparse6}. Under the condition of strong external noise, suprathreshold spiking neurons discharge irregular firings as Geiger counters, and then the population state becomes unsynchronized. However, as the inhibitory recurrent feedback becomes sufficiently strong, this asynchronous state may be destabilized, and then a synchronous population state with stochastic and intermittent individual discharges emerges. Thus, under the balance between strong external excitation and strong recurrent inhibition, fast sparse synchronization was found to occur in both random networks \cite{Sparse1,Sparse2,Sparse3,Sparse4} and globally-coupled networks \cite{Sparse5,Sparse6}.

In brain networks, architecture of synaptic connections has been found to have complex topology (e.g., small-worldness and scale-freeness) which is neither regular nor completely random \cite{Sporns,Buz2,CN1,CN2,CN3,CN4,CN5,CN6,CN7}. In our recent work \cite{Kim}, as a complex network we employed the Watts-Strogatz model for small-world networks which interpolates between regular lattice with high clustering and random graph with short path length via rewiring \cite{SWN1,SWN2,SWN3}. The Watts-Strogatz model may be regarded as a cluster-friendly extension of the random network by reconciling the six degrees of separation (small-worldness) \cite{SDS1,SDS2} with the circle of friends (clustering). We investigated the effect of small-world connectivity on emergence of fast sparsely synchronized rhythms by varying the rewiring probability from short-range to long-range connection \cite{Kim}. When passing a small critical value of the rewiring parameter, fast sparsely synchronized population rhythms were found to emerge in small-world networks with predominantly local connections and rare long-range connections. We note that these small-world networks as well as random graphs are statistically homogeneous because their degree distributions show bell-shaped ones. However, brain networks have been found to show  power-law degree distributions (i.e., scale-free property) in the rat hippocampal networks \cite{SF1,SF2,SF3,SF4} and the human cortical functional network \cite{SF5}. Moreover, robustness against simulated lesions of
mammalian cortical anatomical networks \cite{SF6,SF7,SF8,SF9,SF10,SF11} has also been found to be most similar to that of a scale-free network (SFN) \cite{SF12}. This type of SFNs are inhomogeneous ones with a few ``hubs'' (superconnected nodes), in contrast to statistically homogeneous networks such as random graphs and small-world networks \cite{BA1,BA2}. Many recent works on various subjects of neurodynamics have been done in SFNs with a few percent of hub neurons with an exceptionally large number of connections \cite{SF13,SF14,SF15,SF16}.

The main purpose of our study is to extend previous works on sparse synchronization in statistically homogeneous networks \cite{Sparse1,Sparse2,Sparse3,Sparse4,Sparse5,Sparse6,Kim} to the case of inhomogeneous SFNs with a few  superconnected hubs. We first consider a directed version of the Barab\'{a}si-Albert SFN model with symmetric preferential attachment with the same in- and out-degrees ($l_\alpha^{(in)} = l_\alpha^{(out)}\equiv l_\alpha$).
\cite{BA1,BA2,Bollobas}, and study emergence of sparsely synchronized rhythms by varying $J$ (synaptic inhibition strength) and $D$ (noise intensity) for a fixed attachment degree $l_\alpha$ in an inhibitory population of fast spiking (FS) Izhikevich interneurons \cite{Izhi1,Izhi2,Izhi3,Izhi4}. Fast sparsely synchronized rhythms are found to appear for large values of $J$ and $D$. For a sufficiently large fixed value of $J$, we make an intensive investigation of the population states by increasing $D$. For small $D$, full synchronization with the same population-rhythm frequency $f_p$ and mean firing rate (MFR) $f_i$ of individual neurons occurs. For this case, all the individual neurons exhibit the same behavior, independently of inhomogeneous network structure. As $D$ passes a lower threshold $D_{th,l}$, a transition to partial synchronization with $f_p > {\langle f_i \rangle}$ ($\langle f_i \rangle$: ensemble-averaged MFR) appears due to intermittent discharge of individual neurons. With increasing from $D_{th,l}$, difference between $f_p$ and ${\langle f_i \rangle}$ increases, and sparse synchronization with $f_p > 4 {\langle f_i \rangle}$ emerges when passing a higher threshold $D_{th,h}$. For the case of partial and sparse synchronization, MFRs of individual neurons vary depending on their degrees. As $D$ is further increased and eventually passes a critical value $D^*$, a transition to unsynchronization occurs due to destructive role of noise to spoil the pacing between sparse spikes. The critical value $D^*$ for the transition to unsynchronization is determined by employing a realistic ``thermodynamic'' order parameter, based on the instantaneous population spike rates (IPSR) \cite{RM}. It is also shown that for $D<D^*$, population synchronization emerges in the whole population because the spatial correlation length between the neuronal pairs covers the whole system. Furthermore, the degree of the population synchronization is also measured in terms of two types of realistic ``statistical-mechanical'' measures, based on (1) the occupation and the pacing degrees of the spikes and (2) the  correlations between the IPSR and the instantaneous individual spike rates \cite{RM,Kim-M}. Only for the partial and sparse synchronization, contributions of individual neurons to population synchronization change depending on their degrees, unlike the case of full synchronization. Consequently, individual neuronal dynamics reveal the inhomogeneous network structure for the case of partial and sparse synchronization, which is in contrast to the case of statistically homogeneous random graphs and small-world networks. As a next step, we also investigate the effect of network architecture on sparse synchronization for fixed values of $J$ and $D$ in the following three cases: (1) variation in the degree of symmetric attachment (2) asymmetric preferential attachment of new nodes with different in- and out-degrees (3) preferential attachment between pre-existing nodes (without addition of new nodes). As the degree $l_\alpha$ of symmetric preferential attachment in the first case of network architecture is increased, both the average path length $L_p$ and the betweenness centralization $C_b$ decrease, which results in increased efficiency of communication between nodes. Consequently, the degree of sparse synchronization becomes higher. On the other hand, with increasing $l_\alpha$ the axon ``wire length'' of the network also increases. At an optimal degree $l_\alpha^*$, there is a trade-off between the population synchronization and the wiring economy, and consequently an optimal fast sparsely-synchronized rhythm is found to emerge at a minimal wiring cost in an economic SFN. As the second case of network architecture, we consider an asymmetric preferential attachment of new nodes with different in- and out-degrees ($l_\alpha^{(in)} \neq l_\alpha^{(out)}$). For this asymmetric case, we also measure $L_p$ and $C_b$ by varying the ``asymmetry'' parameter $\Delta l_\alpha$ denoting the deviation from the above symmetric case, and examine how sparse synchronization varies. As the magnitude $| \Delta l_\alpha |$ of asymmetry parameter is increased, both $L_p$ and $C_b$ increase, which leads to decrease in efficiency of communication between nodes. As a result, the degree of sparse synchronization decreases. For both cases of the positive and the negative asymmetries with the same magnitude (e.g., $\Delta l_\alpha$=15 and -15), their values of $L_p$ and $C_b$ are nearly the same because both the inward and the outward edges are equally involved in computation of $L_p$ and $C_b$. However, their synchronization degrees become different because of their distinctly different in-degree distributions affecting individual MFRs. In addition to the above process where preferential attachment is made to newly added nodes with probability $\alpha$, as the third case of network architecture we also consider another process where preferential attachment between pre-existing nodes (without addition of new nodes) is made with probability $\beta$ ($\alpha + \beta =1$). By varying $\beta$, we also measure $L_p$ and $C_b$ and investigate the effect of this $\beta$-process on sparse synchronization. As $\beta$ is increased, communication between pre-existing neurons becomes more efficient due to decrease in both $L_p$ and $C_b$, and hence the degree of sparse synchronization increases. For these three cases of network architecture, dynamics of individual neurons reveal the inhomogeneous structure of the SFN and hence their contributions to sparse synchronization vary depending on their degrees, in contrast to the case of statistically homogeneous random graphs and small-world networks.

This paper is organized as follows. In Sec.~\ref{sec:SFN}, we describe a directed SFN of inhibitory FS Izhikevich interneurons. In Sec.~\ref{sec:FSS}, we first investigate emergence of sparsely synchronized rhythms in a directed  Barab\'{a}si-Albert SFN, and then the effect of network architecture (such as the degree of symmetric attachment, the asymmetric attachment, and the preferential attachment between pre-existing nodes) on fast sparse synchronization is also studied.  Finally, a summary is given in Section \ref{sec:SUM}.

\section{Scale-Free Network of Inhibitory FS Izhikevich Interneurons}
\label{sec:SFN}
We consider an SFN of $N$ inhibitory interneurons equidistantly placed on a one-dimensional ring of radius $N/ 2 \pi$. Here, we employ a directed variant of the Barab\'{a}si-Albert SFN model, composed of
two independent $\alpha-$ and $\beta-$processes which are performed with probabilities $\alpha$ and $\beta$ ($\alpha + \beta =1$), respectively \cite{BA1,BA2,Bollobas}. The diagrams for these two processes generating an SFN are
shown in Fig.~1. The $\alpha$-process corresponds to a directed version of the Barab\'{a}si-Albert SFN model (i.e. growth and preferential directed attachment). For the $\alpha$-process (occurring with the probability $\alpha$), at each discrete time $t$ a new node is added, and it has $l_{\alpha}^{(in)}$ incoming (afferent) edges and $l_{\alpha}^{(out)}$ outgoing (efferent) edges through preferential attachments with $l_{\alpha}^{(in)}$ (pre-existing) source nodes and $l_{\alpha}^{(out)}$ (pre-existing) target nodes, as shown in Fig.~1(a). The (pre-existing) source and target nodes $i$ (which are connected to the new node) are preferentially chosen depending on their out-degrees $d_i^{(out)}$ and in-degrees $d_i^{(in)}$ according to the attachment probabilities $\Pi_{source}(d_i^{(out)})$ and $\Pi_{target}(d_i^{(in)})$, respectively:
\begin{equation}
\Pi_{source}(d_i^{(out)})=\frac{d_i^{(out)}}{\sum_{j=1}^{N_{t -1}}d_j^{(out)}}\;\; \textrm{and} \;\; \Pi_{target}(d_i^{(in)})=\frac{d_i^{(in)}}{\sum_{j=1}^{N_{t -1}}d_j^{(in)}},
\label{eq:AP}
\end{equation}
where $N_{t-1}$ is the number of nodes at the time step $t-1$.
The cases of $l_{\alpha}^{(in)} = l_{\alpha}^{(out)} (\equiv l_{\alpha})$  and $l_{\alpha}^{(in)} \neq l_{\alpha}^{(out)}$ will be referred to as symmetric and asymmetric preferential attachments, respectively.
For the $\beta$-process (occurring with the probability $\beta$), there is no addition of new nodes (i.e., no growth), and symmetric preferential attachments with the same in- and out-degrees [$l_{\beta}^{(in)} = l_{\beta}^{(out)} (\equiv l_{\beta}$)] are made between $l_{\beta}$ pairs of (pre-existing) source and target nodes which are also preferentially chosen according to the attachment probabilities $\Pi_{source}(d_i^{(out)})$ and $\Pi_{target}(d_i^{(in)})$ of Eq.~(\ref{eq:AP}), respectively, such that self-connections (i.e., loops) and duplicate connections (i.e., multiple edges) are excluded [see Fig.~1(b)]. Through the $\beta$-process, degrees of pre-existing nodes are more intensified. For generation of an SFN with $N$ nodes, we start with the initial network at $t=0$, composed of $N_0=50$ nodes where the node 1 is connected bidirectionally to all the other nodes, but the remaining nodes (except the node 1) are sparsely and randomly connected with a low probability $p=0.1$. Then, the $\alpha-$ and $\beta-$processes are repeated until the total number of nodes becomes $N$. For our initial network, the node 1 will be grown as the hub with the highest degree. However, the results (given in Sec.~\ref{sec:FSS}) are independent of the initial networks.

As an element in our neural system, we choose the FS Izhikevich interneuron model which is not only biologically plausible, but also computationally efficient \cite{Izhi1,Izhi2,Izhi3,Izhi4}.
The population dynamics in our SFN is governed by the following set of ordinary differential equations:
\begin{eqnarray}
C\frac{dv_i}{dt} &=& k (v_i - v_r) (v_i - v_t) - u_i +I_{DC} +D \xi_{i} -I_{syn,i}, \label{eq:CIZA} \\
\frac{du_i}{dt} &=& a \{ U(v_i) - u_i \},  \;\;\; i=1, \cdots, N, \label{eq:CIZB}
\end{eqnarray}
with the auxiliary after-spike resetting:
\begin{equation}
{\rm if~} v_i \geq v_p,~ {\rm then~} v_i \leftarrow c~ {\rm and~} u_i \leftarrow u_i + d, \label{eq:RS}
\end{equation}
where
\begin{eqnarray}
U(v) &=& \left\{ \begin{array}{l} 0 {\rm ~for~} v<v_b \\ b(v - v_b)^3 {\rm ~for~} v \ge v_b \end{array} \right. , \label{eq:CIZC} \\
I_{syn,i} &=& g_{syn,i} (v_i - V_{syn});~g_{syn,i} = \frac{J}{d_i^{(in)}} \sum_{j(\ne i)}^N w_{ij} s_j(t), \label{eq:CIZD}\\
s_j(t) &=& \sum_{f=1}^{F_j} E(t-t_f^{(j)}-\tau_l);~E(t) = \frac{1}{\tau_d - \tau_r} (e^{-t/\tau_d} - e^{-t/\tau_r}) \Theta(t). \label{eq:CIZE}
\end{eqnarray}
Here, the state of the $i$th neuron at a time $t$ is characterized by two state variables: the membrane potential $v_i$ and the recovery current $u_i$. In Eq.~(\ref{eq:CIZA}), $C$ is the membrane capacitance, $v_r$ is the resting membrane potential, and $v_t$ is the instantaneous threshold potential. After the potential reaches its apex (i.e., spike cutoff value) $v_p$, the membrane potential and the recovery variable are reset according to Eq.~(\ref{eq:RS}). The units of the capacitance $C$, the potential $v$, the current $u$, and the time $t$ are pF, mV, pA, and ms, respectively.

Unlike Hodgkin-Huxley-type conductance-based models, the Izhikevich model matches neuronal dynamics by tuning the parameters instead of matching neuronal electrophysiology. The parameters $k$ and $b$ are associated with the neuron's rheobase and input resistance, $a$ is the recovery time constant, $c$ is the after-spike reset value of $v$, and $d$ is the total amount of outward minus inward currents during the spike and affecting the after-spike behavior (i.e., after-spike jump value of $u$). Tuning these parameters, the Izhikevich neuron model may produce 20 of the most prominent neuro-computational features of cortical neurons \cite{Izhi1,Izhi2,Izhi3,Izhi4}. Here, we use the parameter values for the FS interneurons (which do not fire postinhibitory rebound spikes) in the layer 5 Rat visual cortex \cite{Izhi3}; $C=20,~v_r=-55,~v_t=-40,~v_p=25,~v_b=-55,~k=1,~a=0.2,~b=0.025,~c=-45,~d=0.$	

Each Izhikevich interneuron is stimulated by using the common DC current $I_{DC}$ (measured in units of pA) and an independent Gaussian white noise $\xi_i$ [see the 3rd and the 4th terms in Eq.~(\ref{eq:CIZA})] satisfying $\langle \xi_i(t) \rangle =0$ and $\langle \xi_i(t)~\xi_j(t') \rangle = \delta_{ij}~\delta(t-t')$, where $\langle\cdots\rangle$ denotes the ensemble average. The noise $\xi$ is a parametric one that randomly perturbs the strength of the applied current $I_{DC}$, and its intensity is controlled by using the parameter $D$ (measured in units of ${\rm pA \cdot {ms}^{1/2}}$). In the absence of noise (i.e., $D=0$), the Izhikevich interneuron exhibits a jump from a resting state to a spiking state via subcritical Hopf bifurcation for $I_{DC,h}=73.7$ by absorbing an unstable limit cycle born via a fold limit cycle bifurcation for $I_{DC,l}=72.8$. Hence, the Izhikevich interneuron shows type-II excitability because it begins to fire with a non-zero frequency \cite{Ex1,Ex2}. As $I_{DC}$ is increased from $I_{DC,h}$, the mean firing rate $f$ increases monotonically. Throughout this paper, we consider a suprathreshold case of $I_{DC}=1500$, where the membrane potential $v$ oscillates very fast with $f=633$ Hz; for more details, refer to Fig.~1 in \cite{Kim}.

The last term in Eq.~(\ref{eq:CIZA}) represents the synaptic coupling of the network. $I_{syn,i}$ of Eq.~(\ref{eq:CIZD}) represents a synaptic current injected into the $i$th neuron; $g_{syn,i}$ represents the synaptic conductance of the $i$th neuron. The synaptic connectivity is given by the connection weight matrix $W$ (=$\{ w_{ij} \}$) where  $w_{ij}=1$ if the neuron $j$ is presynaptic to the neuron $i$; otherwise, $w_{ij}=0$. Here, the synaptic connection is modeled by using the directed SFN (explained in the above). Then, the in-degree of the $i$th neuron, $d_i^{(in)}$ (i.e., the number of synaptic inputs to the neuron $i$) is given by $d_i^{(in)} = \sum_{j(\ne i)}^N w_{ij}$. The fraction of open synaptic ion channels at time $t$ is denoted by $s(t)$. The time course of $s_j(t)$ of the $j$th neuron is given by a sum of delayed double-exponential functions $E(t-t_f^{(j)}-\tau_l)$ [see Eq.~(\ref{eq:CIZE})], where $\tau_l$ is the synaptic delay, and $t_f^{(j)}$ and $F_j$ are the $f$th spike and the total number of spikes of the $j$th neuron at time $t$, respectively. Here, $E(t)$ [which corresponds to contribution of a presynaptic spike occurring at time $0$ to $s(t)$ in the absence of synaptic delay] is controlled by the two synaptic time constants: synaptic rise time $\tau_r$ and decay time $\tau_d$, and $\Theta(t)$ is the Heaviside step function: $\Theta(t)=1$ for $t \geq 0$ and 0 for $t <0$. For the inhibitory GABAergic synapse (involving the $\rm{GABA_A}$ receptors), $\tau_l=1$ ms, $\tau_r=0.5$ ms, and $\tau_d=5$ ms \cite{Sparse6}. The coupling strength is controlled by the parameter $J$ (measured in units of $\rm \mu S$), and $V_{syn}$ is the synaptic reversal potential. Here, we use $V_{syn}=-80$ mV for the inhibitory synapse.

Numerical integration of Eqs.~(\ref{eq:CIZA})-(\ref{eq:CIZB}) is done using the Heun method \cite{SDE} (with the time step $\Delta t=0.01$ ms). For each realization of the stochastic process, we choose a random initial point $[v_i(0),u_i(0)]$ for the $i$th $(i=1,\dots, N)$ neuron with uniform probability in the range of $v_i(0) \in (-50,-45)$ and $u_i(0) \in (10,15)$.

\section{Emergence of fast sparsely synchronized rhythms in scale-free networks}
\label{sec:FSS}
In this section, we study emergence of sparsely synchronized rhythms with stochastic and intermittent neuronal discharges by varying $J$ (synaptic inhibition strength) and $D$ (noise intensity) in SFNs with a few  superconnected hubs. Fast sparsely synchronized rhythms are thus found to appear for large values of $J$ and $D$ by employing both a thermodynamic order parameter and a spatial correlation function  between neuronal pairs. The degree of population synchronization is also characterized in terms of two statistical-mechanical spiking and correlation measures. For this sparse synchronization, contributions of individual neurons to population synchronization vary depending on their degrees, and hence individual neuronal dynamics reveal the inhomogeneous network structure, in contrast to the case of statistically homogeneous random graphs and small-world networks. Furthermore, we also investigate the effect of network architecture on sparse synchronization for fixed $J$ and $D$ by varying $l_{\alpha}$ (i.e., degree of symmetric preferential attachment) and $\Delta l_{\alpha}$ (i.e., asymmetry parameter representing the deviation from the symmetric case) in the $\alpha$-process of adding new nodes and the probability $\beta$ for the $\beta$-process of preferential attachment between (pre-existing) nodes (without addition of new nodes).

We first study a directed version of the Barab\'{a}si-Albert SFN model with symmetric preferential attachment of $l_{\alpha}^{(in)} = l_{\alpha}^{(out)} \equiv l_{\alpha} =25$, composed of $N$ inhibitory FS Izhikevich interneurons equidistantly placed on a one-dimensional ring of radius $N/ 2 \pi$ \cite{BA1,BA2,Bollobas}. The in-degree $d_i^{(in)}$ and the out-degree $d_i^{(out)}$ of individual neurons $i$ show power-law distributions with the same exponent $\gamma=3.0$ \cite{BA1,BA2}, and the average number of synaptic inputs per neuron $M_{syn}^{(in)}$ ($ = \langle d_i^{(in)} \rangle$; $\langle \cdots \rangle$ denotes an ensemble-average over all neurons) is 50, which is nearly the same as that in the small-world network of Ref.~\cite{Kim}. By changing $J$ and $D$, we investigate occurrence of population synchronized states. In computational neuroscience, an ensemble-averaged global potential $V_G$,
\begin{equation}
 V_G (t) = \frac {1} {N} \sum_{i=1}^{N} v_i(t),
\label{eq:GPOT}
\end{equation}
is often used for describing emergence of population synchronization.  However, to directly obtain $V_G$ in real experiments is very difficult. To overcome this difficulty, instead of $V_G$, we use an experimentally-obtainable
IPSR (instantaneous population spike rate) which is often used as a collective quantity showing population behaviors \citep{W_Review,Sparse1,Sparse2,Sparse3,Sparse4,Sparse5,Sparse6}. The IPSR is obtained from the raster plot of neural spikes which is a collection of spike trains of individual neurons. Such raster plots of spikes, where population spike synchronization may be well visualized, are fundamental data in experimental neuroscience. For the synchronous case, ``stripes'' (composed of spikes and representing population synchronization) are found to be formed in the raster plot. Hence, for a synchronous case, an oscillating IPSR appears, while for an unsynchronized case the IPSR is nearly stationary. To obtain a smooth IPSR, we employ the kernel density estimation (kernel smoother) \cite{Kernel}. Each spike in the raster plot is convoluted (or blurred) with a kernel function $K_h(t)$ to obtain a smooth estimate of IPSR, $R(t)$:
\begin{equation}
R(t) = \frac{1}{N} \sum_{i=1}^{N} \sum_{s=1}^{n_i} K_h (t-t_{s}^{(i)}),
\label{eq:IPSRK}
\end{equation}
where $t_{s}^{(i)}$ is the $s$th spiking time of the $i$th neuron, $n_i$ is the total number of spikes for the $i$th neuron, and we use a Gaussian
kernel function of band width $h$:
\begin{equation}
K_h (t) = \frac{1}{\sqrt{2\pi}h} e^{-t^2 / 2h^2}, ~~~~ -\infty < t < \infty.
\label{eq:Gaussian}
\end{equation}
We first consider the case of $D=0$. For sufficiently small $J$, individual interneurons fire too fast to be synchronized. However, as $J$ is increased from zero, MFRs $f_i$ of individual interneurons decrease, and eventually when $J$ passes a critical value $J^* (\simeq 14)$ a transition to full synchronization with the same population-rhythm frequency $f_p$ and MFR $f_i$ of individual neurons occurs. Figures \ref{fig:SD}(a1) and \ref{fig:SD}(a2) show the raster plot of spikes and the IPSR kernel estimate $R(t)$ for small values of $J=100$ and $D=50$, respectively. Clear stripes are formed in the raster plot, and the corresponding IPSR kernel estimate $R(t)$ exhibits large-amplitude regular oscillation with population frequency $f_p=200$ Hz. For this case, individual interneurons fire regularly with the same MFR $f_i$ which is the same as the population frequency $f_p$, and hence complete full synchronization with $f_i = f_p$ occurs, independently of inhomogeneous network structure. However, for the sparsely synchronized cortical rhythms, $f_p: {\langle f_i \rangle} \sim 4:1$ ($\langle f_i \rangle$: ensemble-averaged MFR of individual neurons), unlike the case of full synchronization \cite{Sparse1,Sparse2,Sparse3,Sparse4}. Hence, when the population frequency is much higher than the MFR rate of individual interneurons ($f_p > 4\, {\langle f_i \rangle}$), the synchronization will be referred to as sparse synchronization. For sufficiently large values of $J$ and $D$, sparse synchronization with $f_p > 4\,{\langle f_i \rangle}$ appears. Figures \ref{fig:SD}(b1) and \ref{fig:SD}(b2) show the raster plot of spikes and the IPSR kernel estimate $R(t)$ for $J=1500$ and $D=450$, respectively. For this case, the population frequency $f_p$ of $R(t)$ is about 147 Hz [see Fig.~\ref{fig:SD}(c1)], while the distribution of MFRs $f_i$ of individual neurons is very broad [see Fig.~\ref{fig:SD}(c2)] and the ensemble-averaged MFR $\langle f_i \rangle$ ($ =36$ Hz) is much less than the population frequency $f_p$. Due to this stochastic and intermittent discharge of individual interneurons, stripes in the raster plot become sparse and smeared. Consequently, the amplitude of $R(t)$ becomes smaller. Figure \ref{fig:SD}(d) shows the overall state diagram in the $J-D$ plane. As $D$ is increased, the full synchronization for $D=0$ evolves, depending on the values of $J$, and eventually desynchronization occurs when passing a critical value $D^*$. Plots of $f_p$ and $\langle f_i \rangle$ versus $D$ are also shown in Figs.~\ref{fig:SD}(e1)-\ref{fig:SD}(e4) for $J=$ 100, 500, 1500, and 2000. For small $J$ [$J^* (\simeq 14) < J < 173]$, the full synchronization for $D=0$ develops directly into an unsynchronized state without any other type of intermediate synchronization stage because $f_p = f_i$ (e.g., see the case of $J=100$). However, for $J>173$, the full synchronization for $D=0$ is developed into partial synchronization with $f_p > {\langle f_i \rangle}$ at some lower threshold value $D_{th,l}$ via pitchfork-like bifurcations (e.g., see the cases of $J=500$ 1500, and 2000). With increasing $J$, the difference between $f_p$ and $f_i$ increases abruptly when passing $D_{th,l}$. For $J>1440$, the partial synchronization also evolves into sparse synchronization with $f_p > 4\, {\langle f_i \rangle}$ as $D$ passes a higher threshold $D_{th,h}$ (e.g., see the cases of $J=1500$ and 2000), and eventually when passing a critical value $D^*$, transition to unsynchronization occurs.

For further understanding, we present explicit examples for $J=1500$ which show how the full synchronization is evolved into an unsynchronized state as $D$ is increased. Figures \ref{fig:FSS}(a1)-\ref{fig:FSS}(a5), \ref{fig:FSS}(b1)-\ref{fig:FSS}(b5), and \ref{fig:FSS}(c1)-\ref{fig:FSS}(c5) show the raster plots, the IPSR kernel estimates $R(t)$, and the inter-spike interval (ISI) histograms for $D=100$, 150, 450, 600, and 800, respectively. For $D<D_{th,l} (\simeq 109)$, full synchronization with $f_p = f_i$ occurs (e.g., see the case of $D=100$). All the individual neurons fire regularly with the same MFR $f_i =67$ Hz, which is well shown
in the ISI histogram with a single peak at the global period $T_G$ ($\simeq 14.9$ ms) of $R(t)$ in Fig.~\ref{fig:FSS}(c1). Consequently, clear stripes are formed in the raster plot of spikes and the IPSR kernel estimate $R(t)$ shows large-amplitude regular oscillation with $f_p=67$ Hz [see Figs.~\ref{fig:FSS}(a1)-\ref{fig:FSS}(b1)]. However, when passing the lower threshold $D_{th,l}$, partial synchronization with $f_p > {\langle f_i \rangle}$ appears.
As an example, consider the case of $D=150$. In contrast to the case of full synchronization, the ISI histogram has multiple peaks appearing at multiples of the period $T_G$ ($\simeq 9.3$ ms) of $R(t)$ [see Fig.~\ref{fig:FSS}(c2)]. Similar skipping phenomena of spikings (characterized with multi-peaked ISI histograms) have also been found in networks of coupled inhibitory neurons in the presence of noise where noise-induced hopping from one cluster to another one occurs \cite{GR}, in single noisy neuron models exhibiting stochastic resonance due to a weak periodic external force \cite{Longtin1,Longtin2}, and in inhibitory networks of coupled subthreshold neurons showing stochastic spiking coherence \cite{Kim1,Kim2,Kim3}. ``Stochastic spike skipping'' in coupled systems is a collective effect because it occurs due to a driving by a coherent ensemble-averaged synaptic current, in contrast to the single case driven by a weak periodic force where stochastic resonance occurs. Due to this stochastic spike skipping, partial occupation occurs in the stripes of the raster plot. Thus, the ensemble-averaged MFR $\langle f_i \rangle$ $(\simeq 46$ Hz) of individual interneurons become less than the population frequency $f_p$ ($\simeq 107$ Hz), which results in occurrence of partial synchronization. In contrast to the full-synchronization case of $D=100$, $\langle f_i \rangle$ is decreased, while $f_p$ is increased. For this case of partial synchronization, the density of stripes in the raster plot becomes lower because smaller fraction of total neurons fire in each stripes, and the stripes become smeared, as shown in Fig.~\ref{fig:FSS}(a2). Thus, both the occupation and the pacing degrees of spikes in the raster plot decrease, and consequently a large decrease in the amplitude of $R(t)$ occurs [see Fig.~\ref{fig:FSS}(b2)]. As $D$ is further increased and passes the higher threshold $D_{th,h}$ $(\simeq 400)$, sparse synchronization with $f_p > 4\, {\langle f_i \rangle}$ appears (e.g., see the cases of $D=450$ and 600). The interval between stripes in the raster plot becomes smaller [see Figs.~\ref{fig:FSS}(a3)-\ref{fig:FSS}(a4)], and hence the population frequency of $R(t)$ increases (see Figs.~\ref{fig:FSS}(b3)-\ref{fig:FSS}(b4); $f_p$= 147 and 154 Hz for $D=450$ and 600, respectively). On the other hand, the ensemble-averaged MFR $\langle f_i \rangle$ ($\simeq $36 Hz) for both cases of $D=450$ and 600 is a little decreased in comparison to the case of $D=150$, which results in decrease in density of stripes. We also note that multiple peaks in the ISI histogram overlap and the height of the 1st peak increases, as shown in Figs.~\ref{fig:FSS}(c3)-\ref{fig:FSS}(c4), and hence the stripes become more and more smeared. In this way, both the occupation and the pacing degrees of spikes (seen in the raster plot) decrease. Eventually, when passing the critical value $D^*$ $(\simeq 759$), a transition to unsynchronization occurs. As an example of unsynchronized state, consider the case of $D=800$. Multiple peaks in the ISI histogram become overlapped completely [see Fig.~\ref{fig:FSS}(c5)], and hence spikes in the raster plot are completely scattered, as shown in Fig.~\ref{fig:FSS}(a5). Consequently, the IPSR kernel estimate $R(t)$ in Fig.~\ref{fig:FSS}(b5) becomes nearly stationary (i.e., no population rhythm appears).

In addition to the population dynamics shown in Fig.~\ref{fig:FSS}, we also investigate the dynamics of individual neurons for $J=1500$ to examine whether individual dynamics reveals the inhomogeneous structure of the SFN.
Figures \ref{fig:ID}(a1)-\ref{fig:ID}(a4) show the time-series of membrane potentials $v_i$ of the hub neuron ($i=1$ with the highest degree) and the fastest and slowest peripheral neurons with low degrees ($i:$ varying depending on $D$). For the full-synchronization case of $D=100$, all the individual neurons fire regularly with the same MFR $f_i$ $(\simeq 67$ Hz), as shown in Fig.~\ref{fig:ID}(b1), and hence complete full synchronization occurs, irrespectively of inhomogeneous structure of the SFN. However, for the partial and the sparse synchronization, MFRs vary depending on their degrees. For the partial synchronization of $D=150$, the MFR $f_1$ of the hub neuron $(i=1)$ with highest degree is 32 Hz [which is a little less than the ensemble-averaged MFR $\langle f_i \rangle$ $(\simeq 46$ Hz)], while the MFRs $f_{691}$ and $f_{730}$ of the fastest ($i=691$) and the slowest ($i=730$) peripheral neurons are 87 and 18 Hz, respectively. Hence, MFRs of peripheral neurons with low degrees are distributed broadly around (i.e., above and below) the ensemble-averaged MFR $\langle f_i \rangle$ [denoted by the gray line in Fig.~\ref{fig:ID}(b2)], while MFRs of most of hub neurons with high degrees are less than $\langle f_i \rangle$. As $D$ is further increased and passes the higher threshold $D_{th,h}$, sparse synchronization $f_p > 4\, {\langle f_i \rangle}$ appears. For the sparse-synchronization cases of $D=450$ and 600, distributions of MFRs of individual neurons become more broad when compared with that for the partial-synchronization case of $D=150$, as shown in Figs.~\ref{fig:ID}(b3)-\ref{fig:ID}(b4). The ensemble-averaged MFR $\langle f_i \rangle$ for both cases of $D=450$ and 600 is 36 Hz which is less than that for $D=150$ because more fraction of neurons have lower MFRs for the case of sparse synchronization. Difference in the MFRs of the hub neuron $(i=1)$ and the fastest and the slowest peripheral neurons can also be easily understood in terms of the time-averaged synaptic conductance $\overline{g_{syn,i}}$ of Eq.~(\ref{eq:CIZD}). The synaptic conductance $g_{syn,i}$ of the neuron $i$ is determined mainly by MFRs of pre-synaptic neurons because the fraction of open synaptic ion channels is controlled through the double-exponential function of spikes of pre-synaptic neurons [see Eq.~(\ref{eq:CIZE})]. If the MFR of a pre-synaptic neuron is fast (slow), then its contribution to $g_{syn,i}$ becomes larger (smaller), and hence more (less) inhibition can be given to the post-synaptic neuron. Consequently, the MFR of the post-synaptic neuron becomes slow (fast). Figures \ref{fig:ID}(c1)-\ref{fig:ID}(c4) show the distributions of MFRs of pre-synaptic neurons for the three cases of the hub neuron with $i=1$ (gray region) and the fastest (solid line) and the slowest (dotted line) peripheral neurons. For the case of full synchronization ($D=100$), all pre-synaptic neurons have the same MFR $f_i$ ($\simeq 67$ Hz), irrespectively of degrees of neurons. However, for the partial and sparse synchronization, the distribution of MFRs of pre-synaptic neurons vary depending on post-synaptic neurons. The fastest peripheral neuron has more fraction of pre-synaptic neurons with slower MFRs (as shown by the solid lines) than the hub neuron (gray region), and hence its time-averaged synaptic conductance becomes less than that of the hub neuron. Consequently, its MFR becomes faster than that of the hub neuron. On the other hand, the slowest peripheral neurons have more fraction of pre-synaptic neurons with faster MFRs (as shown by dotted lines) than the hub neuron (gray region), and hence its time-averaged synaptic conductance becomes more than that of the hub neuron. As a result, its MFR becomes slower than that of the hub neuron. In this way, for the partial and sparse synchronization, individual neuronal dynamics vary depending on their degrees, and reveal the inhomogeneous network structure.

As is well known, a conventional order parameter, based on the ensemble-averaged global potential $V_G$, is often used for describing transition from asynchrony to synchrony in computational neuroscience \cite{Order1,Order2,Order3}. Recently, instead of $V_G$, we used an experimentally-obtainable IPSR kernel estimate $R(t)$, and developed a realistic order parameter, which may be applicable in both the computational and the experimental neuroscience \cite{RM,Kim}. The mean square deviation of $R(t)$,
\begin{equation}
{\cal{O}} \equiv \overline{(R(t) - \overline{R(t)})^2},
 \label{eq:Order}
\end{equation}
plays the role of an order parameter $\cal{O}$. (Here the overbar represents the time average.) The order parameter may be regarded as a thermodynamic measure because it concerns just the macroscopic IPSR kernel estimate $R(t)$ without any consideration between $R(t)$ and microscopic individual spikes. In the thermodynamic limit of $N \rightarrow \infty$, the order parameter $\cal{O}$ approaches a non-zero (zero) limit value for the synchronized (unsynchronized) state. Figure \ref{fig:Order}(a) shows a plot of the order parameter versus the noise intensity $D$. For $D < D^*$ $(\simeq 759$), synchronized states exist because the order parameter $\cal{O}$ become saturated to a non-zero limit value for $N \geq 3 \cdot 10^3$. As $D$ passes the critical value $D^*$, a transition to unsynchronization occurs because the values of $\cal {O}$ tends to zero as $N \rightarrow \infty$.
Here we present two explicit examples for the synchronized and the unsynchronized states. First, we consider the population state for $D=700$. As shown in Fig.~\ref{fig:Order}(b1) for $N=10^3$, the raster plot shows sparse stripes of spikes, and $R(t)$ shows a regular oscillation, although there are some variations in the amplitudes. As $N$ is increased to $N=10^4$, stripes in the raster plot become a little more clear, and $R(t)$ also shows a little more regular oscillation [see Fig.~\ref{fig:Order}(b2)]. Consequently, the population state for $D=700$ seems to be synchronized because $R(t)$ tends to show regular oscillations as $N$ goes to the infinity.
As a second example, we consider an unsynchronized case of $D=800$. For $N=10^3$, sparse spikes are scattered without forming any stripes in the raster plot, and $R(t)$ exhibits noisy fluctuations with small amplitude. As $N$ is increased to $10^4$, sparse spikes become more scattered, and consequently $R(t)$ becomes nearly stationary, as shown in Fig.~\ref{fig:Order}(c2). Hence the population state for $D=800$ seems to be unsynchronized because $R(t)$ tends to be nearly stationary as $N$ increases to the infinity.

We further understand the above synchronization-unsynchronization transition in terms of the ``microscopic'' dynamical cross-correlations between neuronal pairs \cite{Kim}. For obtaining dynamical pair cross-correlations, each spike train of the $i$th neuron is convoluted with the Gaussian kernel function $K_h(t)$ of band width $h$ to get a smooth estimate of instantaneous individual spike rate (IISR) $r_i(t)$:
\begin{equation}
r_i(t) = \sum_{s=1}^{n_i} K_h (t-t_{s}^{(i)}),
\label{eq:IISR}
\end{equation}
where $t_{s}^{(i)}$ is the $s$th spiking time of the $i$th neuron, $n_i$ is the total number of spikes for the $i$th neuron, and $K_h(t)$ is given in Eq.~(\ref{eq:Gaussian}). Then, the normalized temporal cross-correlation function $C_{i,j}(\tau)$ between the IISRs $r_i(t)$ and $r_j(t)$ of the $(i,j)$ neuronal pair is given by:
\begin{equation}
C_{i,j}(\tau) = \frac{\overline{\Delta r_i(t+\tau) \Delta r_j(t)}}{\sqrt{\overline{\Delta {r_i}^2(t)}}\sqrt{\overline{\Delta {r_j}^2(t)}}},
\end{equation}
where $\Delta r_i(t) = r_i(t) - \overline{r_i(t)}$ and the overline denotes the time average.
Then, the spatial cross-correlation $C_L$ ($L=1,...,N/2)$ between neuronal pairs separated by a spatial distance $L$ is given by the average of all the temporal cross-correlations between $r_i(t)$ and $r_{i+L}(t)$ $(i=1,...,N)$ at the zero-time lag \cite{Kim}:
\begin{equation}
C_L = \frac{1}{L} \sum_{i=1}^{N} C_{i, i+L}(0) ~~~~ {\rm for~} L=1, \cdots, N/2.
\label{eq:SCC}
\end{equation}
Figure \ref{fig:SC}(a1) shows the plot of the spatial cross-correlation function $C_L$ versus $L$ for $N=10^3$ in the case of full synchronization for $D=100$. The spatial cross-correlation function $C_L$ is nearly non-zero constant $(\simeq 0.97)$ in the whole range of $L$, and hence the correlation length $\eta$ becomes $N/2$ (=500) covering the whole network (note that the maximal distance between neurons is $N/2$ because of the ring architecture on which neurons exist). Consequently, the whole network is composed of just one single synchronized block. For $N=10^4$, the flatness of $C_L$ in Fig.~\ref{fig:SC}(b1) also extends to the whole range ($L=N/2=5000$) of the network, and the correlation length becomes $\eta=5000$, which also covers the whole network. For this case of $D=100$, due to constructive role of noise favoring the pacing between sparse spikes, the correlation length $\eta$ seems to cover the whole network, independently of $N$. Then, the normalized correlation length $\tilde{\eta}$ ($= \frac {\eta} {N}$), representing the ratio of the correlation length $\eta$ to the network size $N$ (i.e., denoting the relative size of synchronized blocks when compared to the whole network size), has a non-zero limit value, $1/2$, and consequently full synchronization emerges in the whole network. However, as $D$ is further increased, the full synchronization breaks up due to stochastic and intermittent discharges of individual neurons, and then partial and sparse synchronization appears. For the cases of partial synchronization ($D=150)$ and sparse synchronization ($D=450$ and 600), plots of $C_L$ are shown in Figs.~\ref{fig:SC}(a2)-\ref{fig:SC}(a4) for $N=10^3$ and in Figs.~\ref{fig:SC}(b2)-\ref{fig:SC}(b4) for $N=10^4$. The values of $C_L$ are also nearly non-zero constants in the whole range of $L$, independently of $N$. Hence, the partial and sparse synchronization appears because the correlation length $\eta$ covers the whole network. The degree of population synchronization may be measured in terms of the average spatial cross-correlation degree $\langle C_L \rangle_L$ given by averaging of $C_L$ over all lengths $L$. Figure \ref{fig:SC}(c) shows the plot of $\langle C_L \rangle_L$ versus $D$. Just after break-up of the full synchronization, $\langle C_L \rangle_L$ drops abruptly, and then decreases slowly to zero. In contrast to the case of population synchronization, the spatial cross-correlation functions $C_L$ for $D=800$ and 1000 are nearly zero for both cases of $N=10^3$ and $10^4$, as shown in Figs.~\ref{fig:SC}(d1)-\ref{fig:SC}(d2) and Figs.~\ref{fig:SC}(e1)-\ref{fig:SC}(e2). For theses cases, due to a destructive role of noise spoiling the pacing between sparse spikes, the correlation lengths $\eta$ become nearly zero, independently of $N$, and hence no synchronization occurs in the network.

By changing $D$ in the whole range of population synchronization, we also measure the degree of population synchronization in terms of a realistic statistical-mechanical spiking measure $M_s$ which was developed in our recent work \cite{RM}. As shown in Figs.~\ref{fig:FSS}(a1)-\ref{fig:FSS}(a4), population spike synchronization may be well visualized in a raster plot of spikes. For a synchronized case, the raster plot is composed of stripes (indicating population synchronization), and the density and the smearing of these stripes represent the degree of the population synchronization. To measure the degree of the population synchronization seen in the raster plot, a statistical-mechanical spiking measure $M_s$, based on $R(t)$, was introduced by considering the occupation pattern and the pacing pattern of the spikes in the stripes \cite{RM}. The spiking measure $M_i$ of the $i$th stripe is defined by the product of the occupation degree $O_i$ of spikes (representing the density of the $i$th stripe) and the pacing degree $P_i$ of spikes (denoting the smearing of the $i$th stripe):
\begin{equation}
M_i = O_i \cdot P_i.
\label{eq:SM1}
\end{equation}
The occupation degree $O_i$ in the $i$th stripe is given by the fraction of spiking neurons:
\begin{equation}
O_i = \frac {N_i^{(s)}} {N},
\label{eq:O}
\end{equation}
where $N_i^{(s)}$ is the number of spiking neurons in the $i$th stripe. For sparse synchronization, $O_i \ll 1$, while $O_i=1$ for full synchronization. The pacing degree $P_i$ of each microscopic spike in the $i$th stripe can be determined in a statistical-mechanical way by taking into account its contribution to the macroscopic IPSR kernel estimate $R(t)$. Each global cycle of $R(t)$ begins from its left minimum, passes the central maximum, and ends at the right minimum; the central maxima coincide with centers of stripes in the raster plot [see Figs.~\ref{fig:FSS}(a1)-\ref{fig:FSS}(a4) and Figs.~\ref{fig:FSS}(b1)-\ref{fig:FSS}(b4)]. An instantaneous global phase $\Phi(t)$ of $R(t)$ is introduced via linear interpolation in the two successive subregions forming a global cycle \cite{RM,GP}; for more details, refer to Fig.~4 in \cite{RM}. The global phase $\Phi(t)$ between the left minimum (corresponding to the beginning point of the $i$th global cycle) and the central maximum is given by
\begin{equation}
\Phi(t) = 2\pi(i-3/2) + \pi \left(
\frac{t-t_i^{(min)}}{t_i^{(max)}-t_i^{(min)}} \right)
 {\rm~~ for~} ~t_i^{(min)} \leq  t < t_i^{(max)}
~~(i=1,2,3,\dots),
\end{equation}
and $\Phi(t)$ between the central maximum and the right minimum (corresponding to the beginning point of the $(i+1)$th cycle) is given by
\begin{equation}
\Phi(t) = 2\pi(i-1) + \pi \left(
\frac{t-t_i^{(max)}}{t_{i+1}^{(min)}-t_i^{(max)}} \right)
 {\rm~~ for~} ~t_i^{(max)} \leq  t < t_{i+1}^{(min)}
~~(i=1,2,3,\dots),
\end{equation}
where $t_i^{(min)}$ is the beginning time of the $i$th global cycle (i.e., the time at which the left minimum of $R(t)$ appears in the $i$th global cycle) and $t_i^{(max)}$ is the time at which the maximum of $R(t)$ appears in the $i$th global cycle. Then, the contribution of the $k$th microscopic spike in the $i$th stripe occurring at the time $t_k^{(s)}$ to $R(t)$ is given by $\cos \Phi_k$, where $\Phi_k$ is the global phase at the $k$th spiking time [i.e., $\Phi_k \equiv \Phi(t_k^{(s)})$]. A microscopic spike makes the most constructive (in-phase) contribution to $R(t)$ when the corresponding global phase $\Phi_k$ is $2 \pi n$ ($n=0,1,2, \dots$), while it makes the most destructive (anti-phase) contribution to $R(t)$ when $\Phi_i$ is $2 \pi (n-1/2)$. By averaging the contributions of all microscopic spikes in the $i$th stripe to $R(t)$, we obtain the pacing degree of spikes in the $i$th stripe:
\begin{equation}
 P_i ={ \frac {1} {S_i}} \sum_{k=1}^{S_i} \cos \Phi_k,
\label{eq:P}
\end{equation}
where $S_i$ is the total number of microscopic spikes in the $i$th stripe.
By averaging $M_i$ of Eq.~(\ref{eq:SM1}) over a sufficiently large number $N_s$ of stripes, we obtain the statistical-mechanical spiking measure $M_s$:
\begin{equation}
M_s =  {\frac {1} {N_s}} \sum_{i=1}^{N_s} M_i.
\label{eq:SM2}
\end{equation}
By varying $D$, we follow $3 \times 10^3$ stripes for each $D$ and characterize population synchronization in terms of $\langle O_i\rangle$ (average occupation degree), $\langle P_i \rangle$ (average pacing degree), and the statistical-mechanical spiking measure $M_s$ for 11 values of $D$ in the synchronized region, and the results are shown in Figs.~\ref{fig:SM}(a)-\ref{fig:SM}(c). In the case of full synchronization for $D < D_{th,l}$, $\langle O_i \rangle$=1 and ${\langle P_i \rangle} \simeq 1$, which results in $M_s \simeq 1$. However, just after break-up of the full synchronization, the average occupation degree $\langle O_i \rangle$ drops abruptly, because of the partial occupation due to stochastic spike skipping, and then it saturates to a non-zero limit value ($\simeq 0.23$). For the case of partial and sparse synchronization, the average pacing degree $\langle P_i \rangle$ also decreases monotonically to zero. Consequently, the statistical-mechanical spiking measure $M_s$ abruptly drops after break-up of the full synchronization, and then slowly decreases to zero, which is similar to the case of the average spatial cross-correlation degree $\langle C_L \rangle_L$ shown in Fig.~\ref{fig:SC}(c). In addition to the spiking measure $M_s$, we also characterize the population synchronization in terms of another statistical-mechanical correlation measure $M_c$, based on the cross-correlations between the IPSR $R(t)$ and the IISRs $r_i(t)$ ($i=1, ..., N$) \cite{Kim-M}. This correlation-based measure $M_c$ may also be regarded as a statistical-mechanical measure because it quantifies the average contribution of (microscopic) IISRs to the (macroscopic) IPSR. The normalized cross-correlation function $C_i (\tau)$ between $R(t)$ and $r_i(t)$ is given by
\begin{equation}
C_i(\tau) = \frac{\overline{\Delta R(t+\tau) \Delta r_i(t)}}{\sqrt{\overline{\Delta {R}^2(t)}}\sqrt{\overline{\Delta {r_i}^2(t)}}},
\label{eq:CCFi}
\end{equation}
where $\tau$ is the time lag, $\Delta R(t) = R(t) - \overline{R(t)}$, $\Delta r_i(t) = r_i(t)-\overline{r_i(t)}$, and the overline denotes the time average.
Then, the statistical-mechanical correlation measure $M_c$ is given by the ensemble-average of $C_i(0)$ at the zero-time lag \cite{Kim-M}:
\begin{equation}
M_c = \frac{1}{N} \sum_{i=1}^{N} C_i(0).
\label{eq:CM}
\end{equation}
Figure \ref{fig:SM}(d) shows the plot of $M_c$ versus $D$. $M_c \simeq 1$ for the case of full synchronization. On the other hand, it drops abruptly just after break-up of the full synchronization, and then slowly decreases to zero, which
is similar to the case of $M_s$ shown in Fig.~\ref{fig:SM}(c).

For further understanding of population synchronization in Fig.~\ref{fig:SM}, we also investigate contributions of individual neuronal dynamics to the population synchronization. Similar to the population occupation, pacing, and spiking measures of Eqs.~(\ref{eq:SM1}), (\ref{eq:O}), and (\ref{eq:P}), we introduce a spiking measure $M_s^{(i)}$ of the $i$th neuron by considering the firing and the pacing degrees of the spikes of the $i$th neuron.
The firing degree $F^{(i)}$, representing the degree of participation of the $i$th neuron to the stripes in the raster plot of spikes, is given by:
\begin{equation}
F^{(i)} = \frac{1}{N_s} \sum_{j=1}^{N_s} F_j^{(i)},
\end{equation}
where $N_s$ is the number stripes for averaging and $F_j^{(i)}$ denotes the participation of the $i$th neuron in the $j$th stripe.
If the $i$th neuron fires in the $j$th stripe (i.e., the spike of the $i$th neuron participates in the $j$th stripe), then $F_j^{(i)}=1$; otherwise $F_j^{(i)}=0$.
The pacing degree of the $i$th neuron, denoting the degree of contributions of the spikes of the $i$th neuron to the IPSR $R(t)$, is given by:
\begin{equation}
P^{(i)} = \frac{1}{S^{(i)}} \sum_{k=1}^{S^{(i)}} \cos \Phi(t_k^{(s)}(i)),
\end{equation}
where $t_k^{(s)}(i)$ is the $k$th spiking time of the $i$th neuron ($k=1, ..., S^{(i)}$), $\Phi(t_k^{(s)}(i))$ is the global phase at $t_k^{(s)}(i)$, and $S^{(i)}$ is the total number of spikes of the $i$th neuron.
Then, the spiking measure $M_s^{(i)}$ of the $i$th neuron is given by the product of the firing and pacing degrees of the $i$th neuron:
\begin{equation}
M_s^{(i)} = F^{(i)} \cdot P^{(i)}.
\label{eq:SM3}
\end{equation}
Figures \ref{fig:IM}(a1)-\ref{fig:IM}(c1) show plots of $F^{(i)}$, $P^{(i)}$, and $M_s^{(i)}$ versus the in-degree $d^{(in)}$ in the case of the full synchronization for $D=100$, respectively.
The values of $F^{(i)}(=1)$, $P^{(i)} (\simeq 0.99)$, and $M_s^{(i)} (\simeq 0.99)$ are constants, independently of the in-degrees, and hence contributions of individual neurons to population
synchronization are the same. On the other hand, $F^{(i)}$, $P^{(i)}$, and $M_s^{(i)}$ vary depending on the in-degrees for the partial and sparse synchronization.
The firing degrees $F^{(i)}$ of individual neurons for $D=150$, 450, and 600 are shown in Figs.~\ref{fig:IM}(a2)-\ref{fig:IM}(a4), respectively. Due to stochastic spike skipping of individual neurons, they spread around their ensemble-averaged values $\langle F^{(i)} \rangle$ [denoted by gray lines and corresponding to the average occupation degree $\langle O_i \rangle$ in Fig.~\ref{fig:SM}(a)], as in  the case of MFRs in Figs.~\ref{fig:ID}(b2)-\ref{fig:ID}(b4). Hence, $F^{(i)}$ of individual neurons seems to be correlated with their MFRs. As $D$ is increased, the ensemble-averaged firing degree $\langle F^{(i)} \rangle$ decreases abruptly and then saturates to a lower limit value, similar to the case of $\langle O_i \rangle$ in Fig.~\ref{fig:SM}(a). Distributions of the pacing degree $P^{(i)}$ and the spiking measure $M_s^{(i)}$ of individual neurons also exhibit spreads from their ensemble-averaged values (represented by gray lines), as shown in Figs.~\ref{fig:IM}(b2)-\ref{fig:IM}(b4) and Figs.~\ref{fig:IM}(c2)-\ref{fig:IM}(c4), respectively. With increase in $D$, the ensemble-averaged pacing degree $\langle P^{(i)} \rangle$, corresponding to the average pacing degree $\langle P_i \rangle$ in Fig.~\ref{fig:SM}(b), shows a gradual decrease when compared to the case of $\langle F^{(i)} \rangle$. Consequently, the ensemble-averaged spiking measure $\langle M_s^{(i)} \rangle$, corresponding to the population spiking measure $M_s$ in Fig.~\ref{fig:SM}(c), abruptly drop after break-up of the full synchronization, mainly due to
sudden decrease in the ensemble-averaged firing degree $\langle F^{(i)} \rangle$, and then slowly decreases. With increasing $D$, the relative variances of $F^{(i)}$, $P^{(i)}$, and $M_s^{(i)}$ from their ensemble-averaged values increase. For additional characterization of individual dynamics, we also introduce the correlation measure $M_c^{(i)}$ of the $i$th neuron, defined by the cross-correlation $C_i(0)$ [see Eq.~(\ref{eq:CCFi})] between the IPSR $R(t)$ and the IISR $r_i(t)$ of the $i$th neuron at the zero-time lag. The ``individual'' correlation measure $M_c^{(i)}$ represents the contribution of the $i$th neuron to the ``population'' correlation measure $M_c$ of Eq.~(\ref{eq:CM}). Figures \ref{fig:IM}(d1)-\ref{fig:IM}(d4) show distributions of $M_c^{(i)}$ versus the in-degree $d^{(in)}$ for $D=100$, 150, 450, and 600, respectively. For the case of the full synchronization ($D=100$), $M_c^{(i)}$ is the same independently on the in-degrees, while for the cases of partial ($D=150$) and sparse ($D=450$ and 600) synchronization $M_c^{(i)}$ spreads around the ensemble-average value $\langle M_c^{(i)} \rangle$ (denoted by gray lines). As $D$ is increased, the ensemble-averaged value $\langle M_c^{(i)} \rangle$ decreases, while the relative variance from $\langle M_c^{(i)} \rangle$ increases, like the case of $\langle M_s^{(i)} \rangle$. In this way, for the partial and sparse synchronization, contributions of individual dynamics to population synchronization depend on their degrees, (although ensemble-averages of individual measures such as
$F^{(i)}$, $P^{(i)}$, and $M_s^{(i)}$ give the average occupation degree $\langle O_i \rangle$, pacing degree $\langle P_i \rangle$, and spiking measure $M_s$ in the whole population,) and reveal the inhomogeneous structure of the SFN, in contrast to statistically homogeneous networks such as the random graph and the small-world network.

From now on, we investigate the effect of network architecture on sparse synchronization for fixed values of $J=1500$ and $D=450$ in the following three cases. As the first case of network architecture, we consider the effect of the degree $l_{\alpha}$ of the symmetric preferential attachment ($l_{\alpha}^{(in)}$ = $l_{\alpha}^{(out)} \equiv l_{\alpha})$ on sparse synchronization. Figures~\ref{fig:ESFN}(a1)-\ref{fig:ESFN}(a5) and Figs.~\ref{fig:ESFN}(b1)-\ref{fig:ESFN}(b5) show the raster plots and the IPSR kernel estimates $R(t)$ for $l_{\alpha}=15$, 20, 25, 40, and 45, respectively. For $l_{\alpha}$ is less than a threshold value $l_\alpha^{th} (\simeq 17)$, no population synchronization occurs. As an example of unsynchronization, we consider a case of $l_{\alpha}=15$ where spikes in the raster plot are completely scattered and the IPSR kernel estimate $R(t)$ becomes nearly stationary, as shown in Figs.~\ref{fig:ESFN}(a1) and \ref{fig:ESFN}(b1), respectively. When passing the threshold value $l_\alpha^{th}$, a transition to sparse synchronization occurs. For example, for $l_{\alpha}=20$ stripes appear in the raster plot of spikes, and the IPSR kernel estimate $R(t)$ shows regular oscillation [see Figs.~\ref{fig:ESFN}(a2) and \ref{fig:ESFN}(b2)]. As $l_\alpha$ is further increased, the stripes in the raster plot become more and more dense and clear, and the IPSR kernel estimates $R(t)$ show larger-amplitude regular oscillations, as shown in Figs.~\ref{fig:ESFN}(a3)-\ref{fig:ESFN}(a5) and Figs.~\ref{fig:ESFN}(b3)-\ref{fig:ESFN}(b5), respectively. Hence, as $l_{\alpha}$ is increased, the degree of sparse synchronization becomes better. For characterization of the effect of $l_\alpha$ on network topology, we also study the local property of the SFN in terms of the in- and out-degrees. Figures \ref{fig:ESFN}(c1)-\ref{fig:ESFN}(c5) show the plots of the out-degree $d^{(out)}$ versus the in-degree $d^{(in)}$ for $l_{\alpha}=15$, 20, 25, 40, and 45, respectively. The in- and out-degrees are distributed nearly symmetrically around the diagonal, and with increasing $l_\alpha$ they are shifted upward because of increase in the in- and out-degrees. Based on these degree distributions, we classify the nodes into the hub group (consisting of the head hub with the highest degree and the secondary hubs with higher degrees) and the peripheral group (composed of a majority of nodes with lower degrees). As an example, we consider the case of $l_\alpha =25$, and explain how to classify the nodes into the hub and the peripheral groups. For this case, the histogram for fraction of nodes versus the in-degree $d^{(in)}$ (which is also similar to that for the case of out-degree $d^{(out)}$) is shown in Fig.~\ref{fig:ESFN}(d). The majority of peripheral nodes have their degrees near the peak at $d^{(in)}=25$, while the minority of hubs have their degrees in the long-tail part. For convenience, we choose the threshold $d_{th}^{(in)}$ for the in-degree (denoted by the vertical dotted line in Fig.~\ref{fig:ESFN}(d) and separating the hub and the peripheral groups) whose fraction of nodes is $0.002$ (i.e., $0.2 \%$). Similarly, we also choose the threshold $d_{th}^{(out)}$ for the out-degree, which is the same as $d_{th}^{(in)}$. (Hereafter, we choose the thresholds $d_{th}^{(in)}$ and $d_{th}^{(out)}$ for both the in- and out-degrees whose fractions of nodes are $0.2 \%$). For visualization, the peripheral group is enclosed by rectangles (determined by both thresholds $d_{th}^{(in)}$ and $d_{th}^{(out)}$) in Figs.~\ref{fig:ESFN}(c1)-\ref{fig:ESFN}(c5). The hub group (outside the rectangle) is composed of about 100 nodes (i.e., approximately $10 \%$ of the total neurons), where the node 1 (denoted by the open circle) corresponds to the head hub with the highest degree and the other ones will be called as secondary hubs. This kind of degree distribution is a ``comet-shaped'' one; the peripheral and the hub groups correspond to the coma (surrounding the nucleus) and the tail of the comet, respectively. In addition to the in- and out-degrees of individual nodes, we study the group properties of the SFN in terms of the average path length $L_p$ and the betweenness centralization $C_b$ by varying $l_{\alpha}$. The average path length $L_p$, denoting typical separation between two nodes in the network, is given by the average of the shortest path lengths of all neuronal pairs:
\begin{equation}
L_p = \frac{1}{N(N-1)} \sum_{i=1}^{N}\sum_{j=1 (j \ne i)}^{N} l_{ij},
\end{equation}
where $l_{ij}$ is the shortest path length from the node $i$ to the node $j$. We note that $L_p$ characterizes the global efficiency of information transfer between distant nodes. In the network science, centrality refers to indicators which identify the most important nodes within the network (i.e., the centrality indices are answers to the question ``which nodes are most central?''). Historically first and conceptually simplest is the degree centrality (explained above), which is defined by the number of edges of a node. This degree centrality represents the potentiality in communication activity. Superconnected hubs participate in the mainstream of information flow in the network, while peripheral nodes with a few links makes no active participation in the communication process. Betweenness is also another centrality measure of a node within the network. Betweenness centrality of the node $i$ denotes the fraction of all the shortest paths between any two other nodes that pass through the node $i$ \cite{BC1,BC2}:
\begin{equation}
B_i = \sum_{j=1(j \ne i)}^{N} \sum_{k=1(k \ne j ~ \& ~ k \ne i)}^{N} \frac{\sigma_{jk}(i)}{\sigma_{jk}},
\end{equation}
where $\sigma_{jk}(i)$ is the number of shortest paths from the node $j$ to the node $k$ passing through the node $i$ and $\sigma_{jk}$ is the total number of shortest paths from the node $j$ to the node $k$. This betweenness centrality $B_i$ characterizes the potentiality in controlling communication between other nodes in the rest of the network. In our SFN, the head hub (i.e., node 1) with the highest degree is also found to have the maximum betweenness centrality $B_{max}$, and hence the head hub has the largest load of communication traffic passing through it. To examine how evenly the betweenness centrality is distributed among nodes (i.e., how evenly the load of communication traffic is distributed among nodes), we consider the group betweenness centralization, representing the degree to which the maximum betweenness centrality $B_{max}$ of the head hub exceeds the betweenness centrality of all the other nodes. The betweenness centralization $C_b$ is given by the sum of differences between the maximum betweenness centrality $B_{max}$ of the head hub and the betweenness centrality $B_i$ of other node $i$, and normalized by dividing the sum of differences with its maximum possible value \cite{BC1,BC2}:
\begin{equation}
C_b=\frac{\sum_{i=1}^{N} (B_{max} - B_i)}{{\rm max}{\sum_{i=1}^{N} (B_{max} - B_i)}};~ {\rm max}{\sum_{i=1}^{N} (B_{max} - B_i)}= \frac {(N-1)(N^2 - 3 N +2)} {2},
\end{equation}
where the maximum sum of differences in the denominator corresponds to that for the star network.
Large $C_b$ implies that load of communication traffic is concentrated on the head hub, and hence the head hub tends to become overloaded by the communication traffic passing through it. For this case, it becomes difficult to get efficient communication between nodes due to destructive interference between so many signals passing through the head hub \cite{BC3}. Figures \ref{fig:ESFN}(e) and \ref{fig:ESFN}(f) show the plots of the average path length $L_p$ and the betweenness centralization $C_b$ versus $l_{\alpha}$, respectively. With increasing $l_\alpha$ both $L_p$ and $C_b$ decrease monotonically to non-zero values. Decrease in $L_p$ implies reduction in intermediate mediation of nodes controlling the communication in the whole network (i.e., reduction in total centrality $B_{tot}$ given by the sum of centralities of all nodes). How the total betweenness $B_{tot}$ decreases with increase in $l_\alpha$ may be seen explicitly in Fig.~\ref{fig:ESFN}(g). The maximum betweenness $B_{max}$ of the head hub is much more reduced than the average centralities of the secondary hubs and the peripheral nodes, ${\langle B \rangle}_{hub}$ and ${\langle B \rangle}_{peri}$, which leads to decrease in differences between $B_{max}$ of the head hub and $B_i$ of other nodes (i.e., variation between centralities of nodes is reduced). Hence, as the result of increase in $l_\alpha$, typical separation between two nodes in the network becomes shorter and load of communication traffic becomes less concentrated on the head hub (i.e., the load is more evenly distributed among nodes). Consequently, as $l_\alpha$ is increased, efficiency of communication between nodes becomes better, which may result in the increase in the degree of sparse synchronization. The statistical-mechanical spiking measure $M_s$ of Eq.~(\ref{eq:SM2}) for the synchronization degree (denoted by solid circles) is shown in Fig.~\ref{fig:ESFN}(h). As $l_{\alpha}$ is increased, the degree of sparse synchronization increases and tends to become saturated. However, with increasing $l_{\alpha}$, the network axon wiring length becomes longer due to increase in the long-range connections. Longer axonal connections are expensive because of material and energy costs. Hence, in view of dynamical efficiency we search for optimal population rhythm emerging at a minimal wiring cost. We then calculate the wiring length by varying $l_{\alpha}$ on a ring of radius $r$ (=$N/ 2 \pi$) where nodes are placed equidistantly. The axonal wiring length, $L_w^{(ij)}$, between the node $i$ and the node $j$ is given by the arc length between two nodes $i$ and $j$ on the ring:
\begin{equation}
   L_w^{(ij)} = \left\{
\begin{array}{l}
   |j-i| \; \textrm{for} \; |j-i| \le \frac{N}{2}\\
   N-|j-i| \; \textrm{for} \; |j-i| > \frac{N}{2}.
\end{array}
\right.
\end{equation}
Then, the total wiring length is:
\begin{equation}
   L_w^{total} = \sum_{i=1}^{N} \sum_{j=1 (j \ne i)}^{N} a_{ij} \cdot L_w^{(ij)},
\end{equation}
where $a_{ij}$ is the $ij$ element of the adjacency matrix $A$ of the network.
The connection between vertices in the network is represented by its $N \times N$ adjacency matrix $A$ $(=\{a_{ij} \})$ whose element values are $0$ or $1$. If $a_{ij}=1$, then an edge from the vertex $i$ to the vertex $j$ exists; otherwise no such edges exists. This adjacency matrix $A$ corresponds to the transpose of the connection weight matrix $W$ in Sec.~\ref{sec:SFN}. We get a normalized wiring length ${\cal{L}}_w$ by dividing $L_w^{total}$ with $L_{w,global}^{total}$ $[=\sum_{i=1}^{N} \sum_{j=1 (j \ne i)}^{N} L_w^{(ij)}]$ which is the total wiring length for the global-coupled case:
\begin{equation}
{\cal{L}}_w= \frac{L_w^{total}}{L_{w,global}^{total}}.
\end{equation}
Open circles in the Fig.~\ref{fig:ESFN}(h) denote the normalized wiring length ${\cal{L}}_w$. It increases linearly with respect to $l_{\alpha}$. Hence, as $l_{\alpha}$ is increased, the wiring cost becomes expensive. An optimal rhythm may emerge through tradeoff between the synchronization degree $M_s$ and the wiring cost ${\cal{L}}_w$. To this end, a dynamical efficiency $\cal{E}$ is given by \cite{Buz2,Kim}:
\begin{equation}
 {\cal{E}}= \frac{\textrm{Synchronization\, Degree}~ (M_s)}{\textrm{Normalized\, Wiring\, Length}~ ({{\cal{L}}_w})}.
\end{equation}
Figure \ref{fig:ESFN}(i) shows plot of $\cal{E}$ versus $l_\alpha$. For $l_{\alpha}=l_{\alpha}^*$ $(=34)$, an optimal rhythm is found to emerge at a minimal wiring cost in an economic SFN. An optimal fast sparsely synchronized rhythm is shown in Figs.~\ref{fig:ESFN}(j1)-\ref{fig:ESFN}(j2). Sparse stripes appear successively in the raster plot of spikes. Hence, the IPSR kernel estimate $R(t)$ shows a regular oscillation at a population frequency $f_p$ $(\simeq 147$ Hz), while individual neurons fire stochastically and sparsely at the ensemble-averaged MFR $\langle f_i \rangle = 34$ Hz.

So far, we studied the case of symmetric attachment with $l_{\alpha}^{(in)} = l_{\alpha}^{(out)} \equiv l_{\alpha}$. As the second case of network architecture, we consider the case of asymmetric preferential attachment $l_{\alpha}^{(in)} \neq l_{\alpha}^{(out)}$. We set $l_{\alpha}^{(in)}=l_{\alpha}+\Delta l_{\alpha}$ and $l_{\alpha}^{(out)}=l_{\alpha}-\Delta l_{\alpha}$ such that $l_{\alpha}^{(in)} + l_{\alpha}^{(out)} = 2\,l_{\alpha}=$ constant, and investigate the effect of asymmetric attachment on sparse synchronization by varying the asymmetry parameter $\Delta l_{\alpha}$ for $l_{\alpha}=25$. For comparison, the raster plot and the IPSR kernel estimate $R(t)$ for the symmetric case of $\Delta l_\alpha=0$ (i.e., $l_{\alpha}^{(in)} = l_{\alpha}^{(out)}=25$) are shown in Figs.~\ref{fig:AA}(a2) and \ref{fig:AA}(b2), respectively. Figure \ref{fig:AA}(a1) shows the raster plot for the case of negative asymmetric attachment with $\Delta l_{\alpha}=-15$ [i.e., $l_{\alpha}^{(in)}=10$ and $l_{\alpha}^{(out)}=40$]. When compared with the case of the symmetric attachment, the stripes in the raster plot
are much more smeared, while they are a little more dense. In contrast, for the case of positive asymmetric attachment with $\Delta l_{\alpha}=15$ [i.e., $l_{\alpha}^{(in)}=40$ and $l_{\alpha}^{(out)}=10$], the stripes
are less smeared but more sparse in comparison to the case of symmetric attachment, as shown in Fig.~\ref{fig:AA}(a3). The amplitudes of the IPSR kernel estimates $R(t)$ for both cases of $\Delta l_{\alpha}=-15$ and 15 become smaller than that for the symmetric attachment [compare Figs.~\ref{fig:AA}(b1) and \ref{fig:AA}(b3) with Fig.~\ref{fig:AA}(b2)]. When the two asymmetric cases are compared, the amplitude of $R(t)$ for $\Delta l_{\alpha}=15$ is a little larger than that for $\Delta l_{\alpha}=-15$. In this way, the degree of sparse synchronization becomes reduced as the magnitude of the asymmetry parameter $| \Delta l_\alpha |$ is increased. Depending on the sign of the asymmetry parameter $\Delta l_{\alpha}$, the synchronization degree also differs, in spite of the same magnitude of $\Delta l_{\alpha}$ (e.g., $\Delta l_{\alpha}$ = 15 and -15). This difference between the cases of $\Delta l_{\alpha}$ = 15 and -15 occurs due to different in-degree distributions affecting the synaptic inputs to individual neurons [see Eq.~(\ref{eq:CIZD})], which will be explained in Fig.~\ref{fig:IDAA}.
Next, we study the effect of $\Delta l_\alpha$ on the average path length $L_p$ and the betweenness centralization $C_b$. Figures \ref{fig:AA}(c) and \ref{fig:AA}(d) show plots of $L_p$ and $C_b$ versus $\Delta l_{\alpha}$, respectively. Both $L_p$ and $C_b$ increase symmetrically with increasing $| \Delta l_{\alpha} |$, independently of the sign of $\Delta l_{\alpha}$. Since both inward and outward links are involved equally in computation of $L_p$ and $C_b$, the values of $L_p$ and $C_b$ for both cases of different signs but the same magnitude (i.e., $\Delta l_{\alpha}$ and -$\Delta l_{\alpha}$) become the same, unlike the above case of population synchronization where only the inward synaptic inputs affect. As $| \Delta l_{\alpha} |$ is increased, mismatching between the in- and out-degrees of nodes is increased, which leads to increase in $L_p$. This increase in $L_p$ implies enhancement of intermediate mediation of nodes controlling communication in the network (i.e., enhancement in total betweenness $B_{tot}$). As shown in Fig.~\ref{fig:AA}(e), with increasing $| \Delta l_\alpha |$ the maximum betweenness $B_{max}$ of the head hub is much more enhanced than the average centralities of the secondary hubs and the peripheral nodes, ${\langle B \rangle}_{hub}$ and ${\langle B \rangle}_{peri}$, which leads to increase in differences between $B_{max}$ of the head hub and $B_i$ of other nodes (i.e., variation between centralities of nodes is increased). Hence, as $| \Delta l_\alpha |$ is increased, typical separation between two nodes in the network becomes longer and load of communication traffic becomes more concentrated on the head hub. Consequently, with increasing $| \Delta l_\alpha |$, efficiency of communication between nodes becomes worse, which may result in decrease in the degree of sparse synchronization. However, unlike the change in $L_p$ and $C_b$, sparse synchronization varies depending on the sign of $\Delta l_\alpha$. Figures \ref{fig:AA}(f1)-\ref{fig:AA}(f2) show plots of the average occupation degree $\langle O_i \rangle$ and the average pacing degree $\langle P_i \rangle$ versus $\Delta l_{\alpha}$. As $\Delta l_{\alpha}$ is decreased from the symmetric case (i.e., $\Delta l_\alpha =0$), $\langle O_i \rangle$ increases, while it decreases with increasing $\Delta l_{\alpha}$ from 0. On the other hand, with decrease in $\Delta l_{\alpha}$ from 0, $\langle P_i \rangle$ decreases much, while it increases and tends to become saturated with increase in $\Delta l_{\alpha}$ from 0. As a result, the statistical-mechanical spiking measure $M_s$, given by taking into consideration both the occupation and the pacing degrees, has its peak at $\Delta l_\alpha =0$ (i.e., symmetric case), as shown in Fig.~\ref{fig:AA}(f3). Hence, $M_s$ decreases in both positive and negative directions with increasing $| \Delta l_\alpha |$ from 0. The decreasing rate depends on the sign of $\Delta l_\alpha$: $M_s$ for $\Delta l_\alpha <0$ decreases more rapidly than that for $\Delta l_\alpha >0$. For example, $M_s$ for $\Delta l_\alpha=15$ is higher than that for $\Delta l_\alpha=-15$. For more clear presentation, we normalize the occupation degree, the pacing degree, and the spiking measures by dividing them with their ensemble-averaged values for the symmetric case. Then, the normalized occupation degree $\langle \widetilde{O}_i \rangle$, pacing degree $\langle \widetilde{P}_i \rangle$, and spiking measure $\widetilde{M}_s$ are shown in Fig.~\ref{fig:AA}(g). As $\Delta l_\alpha$ is decreased from 0, $\langle \widetilde{O}_i \rangle$ increases, while $\langle \widetilde{P}_i \rangle$ decreases much more, and hence $\widetilde{M}_s$ decreases. On the other hand, as $\Delta l_\alpha$ is increased from 0, $\langle \widetilde{P}_i \rangle$ increases, while $\langle \widetilde{O}i \rangle$ decreases much more, and hence $\widetilde{M}_s$ also decreases. Furthermore, since the variation from the symmetric case is larger for the case of $\Delta l_\alpha <0$, its spiking measure $M_s$ becomes less than that for the positive asymmetric attachment with the same magnitude (e.g., $M_s$ for $\Delta l_\alpha =-15$ is less than that for $\Delta l_\alpha = 15$).

To understand how the sparse synchronization varies differently depending on the sign of the asymmetry parameter $\Delta l_\alpha$, we also investigate contributions of individual neuronal dynamics on the population synchronization. We first consider the effect of $\Delta l_\alpha$ on the degree distribution of nodes. Figures \ref{fig:IDAA}(a1)-\ref{fig:IDAA}(a3) show plots of the out-degree $d^{(out)}$ versus the in-degree $d^{(in)}$ for $\Delta l_\alpha=-15$, 0, and 15, respectively. A majority of peripheral nodes with lower degrees are enclosed by rectangles, while hubs with higher degree lie outside the rectangles. For the case of symmetric attachment (i.e., $\Delta l_\alpha=0$), the in- and out-degrees are distributed nearly symmetrically around the diagonal. Hence, the in-degrees of the hubs and the peripheral nodes are nearly the same as the out-degrees, respectively. On the other hand, the degree distributions vary significantly for the case of asymmetric attachment. For $\Delta l_\alpha=-15$, the in-degrees of peripheral nodes are less than their out-degrees, while the in-degrees of hubs are much more than their out-degrees (i.e., ``popular'' hubs with $d^{(in)} \gg d^{(out)}$ appear). Thus, the distribution of in-degrees is broad, while the distribution of out-degrees is narrow (i.e., the distribution for $\Delta l_\alpha=-15$ seems to be similar to that obtained through clockwise rotation of the symmetric distribution for $\Delta l_\alpha=0$ about a center), as shown in Fig.~\ref{fig:IDAA}(a1). In contrast, the out-degrees of peripheral nodes for $\Delta l_\alpha=15$ are less than their in-degrees, while the out-degrees of hubs are much more than their in-degrees (i.e., ``social'' hubs with $d^{(out)} \gg d^{(in)}$ emerge). Thus, the distribution of in-degrees is narrow, while the distribution of out-degrees is wide (i.e., the distribution for $\Delta l_\alpha=15$ seems to be similar to that obtained through counter-clockwise rotation of the symmetric distribution for $\Delta l_\alpha=0$ about a center), as shown in Fig.~\ref{fig:IDAA}(a3). We note that individual dynamics  vary depending on the synaptic inputs with the in-degree $d^{(in)}$ of Eq.~(\ref{eq:CIZD}). Hence, the in-degree distribution affects the dynamics of individual neurons. Figures \ref{fig:IDAA}(b1)-\ref{fig:IDAA}(b3) show the power-law distributions of in-degrees for $\Delta l_\alpha=-15$, 0, and 15, respectively. As is well known, the exponent for $\Delta l_\alpha=0$ is $\gamma=3.0$ \cite{BA1,BA2}. On the other hand, $\gamma = 2.0$ for $\Delta l_\alpha=-15$ because of broad distribution, while $\gamma = 4.7$ for $\Delta l_\alpha=15$ because of narrow distribution.
Based on these in-degree distributions, we study MFRs of individual neurons. Figures \ref{fig:IDAA}(c1)-\ref{fig:IDAA}(c3) and Figs.~\ref{fig:IDAA}(d1)-\ref{fig:IDAA}(d3) show plots of MFR versus $d^{(in)}$ and histograms for fraction of neurons versus MFR for $\Delta l_\alpha=-15$, 0, and 15, respectively. For the case of symmetric attachment (i.e., $\Delta l_\alpha=0$), the ensemble-averaged MFR $\langle f_i \rangle$ [denoted by the horizontal gray line in Fig.~\ref{fig:IDAA}(c2)] is approximately 36 Hz. Since the in-degree of a peripheral neuron is small, its pre-synaptic neurons belong to a small subset of the whole population. Hence, the MFRs of the peripheral neurons may change depending on the average MFR of pre-synaptic neurons in the small subset. If MFRs of the pre-synaptic neurons (in the small subset) is fast (slow) on average, then the post-synaptic peripheral neuron may receive more (less) synaptic inhibition, and hence its MFR becomes slow (fast). As a result, the MFRs of the peripheral neurons are distributed broadly around the ensemble-averaged gray line. The average MFR ${\langle f_i \rangle}_{peri}$ ($\simeq 38$ Hz) of peripheral neurons is a little faster than the ensemble-averaged MFR $\langle f_i \rangle$ because MFRs of the peripheral neurons are distributed a little more above the horizontal gray line. On the other hand, the pre-synaptic neurons of a hub neuron with higher in-degree belong to a relatively larger subpopulation of the whole network. Since MFRs of the pre-synaptic neurons in the larger subset represent approximately those in the whole population, variation in the synaptic inhibitions received by the hub neurons is small, and hence the distribution of MFRs of the hub neurons becomes narrow. Moreover, since ${\langle f_i \rangle}_{peri} > {\langle f_i \rangle}$, the average MFR ${\langle f_i \rangle}_{hub}$ ($\simeq 25$ Hz) of hub neurons becomes slower than the ensemble-averaged MFR $\langle f_i \rangle$. Thus, MFRs of the hub neurons are narrowly distributed below the ensemble-averaged horizontal gray line. We then consider the case of the asymmetric attachment in comparison with the case of symmetric attachment. For $\Delta l_\alpha = -15$, the in-degrees of peripheral neurons are lower, while those of hub neurons are much higher [compare Figs.~\ref{fig:IDAA}(a1) and \ref{fig:IDAA}(b1) with Figs.~\ref{fig:IDAA}(a2) and \ref{fig:IDAA}(b2)]. Hence, the pre-synaptic neurons of a peripheral neuron belongs to a smaller subpopulation in the whole network. Following the same argument given in the above case of $\Delta l_\alpha=0$, MFRs of the peripheral neurons are distributed around the ensemble-averaged horizontal gray line more broadly than those for $\Delta l_\alpha=0$ [compare Fig.~\ref{fig:IDAA}(c1) with Fig.~\ref{fig:IDAA}(c2)]. As shown in Fig.~\ref{fig:IDAA}(d1), peripheral neurons with faster MFRs appear in comparison to the case of $\Delta l_\alpha=0$ shown in
Fig.~\ref{fig:IDAA}(d2), and hence the average MFR ${\langle f_i \rangle}_{peri}$ ($\simeq 50$ Hz) of peripheral neurons becomes faster than that for $\Delta l_\alpha=0$, which also leads to increase in the ensemble-averaged MFR ${\langle f_i \rangle}$ $(\simeq 47$ Hz) in the whole population, due to the majority of peripheral neurons. On the other hand, due to higher in-degrees, variation in the synaptic inhibitions received by the hub neurons becomes smaller, and hence the distribution of MFRs of hubs becomes more narrow. Furthermore, since ${\langle f_i \rangle}_{peri}$ of peripheral neurons is increased, the average MFR ${\langle f_i \rangle}_{hub}$ ($\simeq 24$ Hz) of hub neurons decreases. Then, the MFRs of the hub neurons are more narrowly distributed much below the ensemble-averaged horizontal gray line [compare Fig.~\ref{fig:IDAA}(c1) with Fig.~\ref{fig:IDAA}(c2)]. We next consider the case of $\Delta l_\alpha = 15$. For this case, the in-degrees of peripheral neurons are increased, while those of hub neurons are much decreased [compare Figs.~\ref{fig:IDAA}(a3) and \ref{fig:IDAA}(b3) with Figs.~\ref{fig:IDAA}(a2) and \ref{fig:IDAA}(b2)], in contrast to the case of $\Delta l_\alpha = -15$. Hence, the pre-synaptic neurons of a peripheral neuron belongs to a little larger subpopulation in the whole network, and hence MFRs of the peripheral neurons are distributed around the ensemble-averaged horizontal gray line much narrowly than those for $\Delta l_\alpha=0$ [compare Fig.~\ref{fig:IDAA}(c3) with Fig.~\ref{fig:IDAA}(c2)]. As shown in Fig.~\ref{fig:IDAA}(d3), peripheral neurons with slower MFRs appear in comparison to the case of $\Delta l_\alpha=0$ shown in Fig.~\ref{fig:IDAA}(d2), and hence the average MFR ${\langle f_i \rangle}_{peri}$ ($\simeq 29$ Hz) of peripheral neurons becomes slower than that for $\Delta l_\alpha=0$, which also leads to decrease in the ensemble-averaged MFR ${\langle f_i \rangle}$ $(\simeq 28$ Hz) in the whole population, because of the majority of peripheral neurons.
Due to this narrow distribution of MFRs of peripheral neurons, variation in the synaptic inhibitions received by the hub neurons also becomes smaller, and hence the distribution of MFRs of hubs also becomes narrow.
Moreover, since ${\langle f_i \rangle}_{peri}$ of peripheral neurons is decreased, the average MFR ${\langle f_i \rangle}_{hub}$ ($\simeq 26$ Hz) of hub neurons increases. Then, the MFRs of the hub neurons are more narrowly distributed just below the ensemble-averaged horizontal gray line [compare Fig.~\ref{fig:IDAA}(c3) with Fig.~\ref{fig:IDAA}(c2)].

Based on the above distributions of MFRs, we study contributions of individual dynamics on the sparse synchronization. Figures \ref{fig:IDAA}(e1)-\ref{fig:IDAA}(e3) show plots of the firing degree $F^{(i)}$ of individual neurons versus the in-degree $d^{(in)}$ for $\Delta l_\alpha=$-15, 0, and 15, respectively. We note that distributions of the firing degree $F^{(i)}$ of individual neurons are strongly correlated with their distributions of MFRs [compare Figs.~\ref{fig:IDAA}(e1)-\ref{fig:IDAA}(e3) with Figs.~\ref{fig:IDAA}(c1)-\ref{fig:IDAA}(c3)]. Similar to the case of MFRs, $F^{(i)}$ spreads around the ensemble-averaged value $\langle F^{(i)} \rangle$ [denoted by gray lines and corresponding to the average occupation degree $\langle O_i \rangle$ in Fig.~\ref{fig:AA}(f1)]. As $\Delta l_\alpha$ is increased, $\langle F^{(i)} \rangle$ decreases, which results in decrease in $\langle O_i \rangle$ in Fig.~\ref{fig:AA}(f1). The variation of $F^{(i)}$ about $\langle F^{(i)} \rangle$ also decreases with increasing $\Delta l_\alpha$. Distributions of the pacing degree $P^{(i)}$ of individual neurons also show spreads from their ensemble-averaged values [represented by gray lines and corresponding to the average pacing degree  in Fig.~\ref{fig:AA}(f2)], as shown in Figs.~\ref{fig:IDAA}(f1)-\ref{fig:IDAA}(f3). As $\Delta l_\alpha$ is increased, both the ensemble-averaged MFR and the variation decrease, and hence the ensemble-averaged pacing degree $\langle P^{(i)} \rangle$ shows an increase, which also leads to increase in $\langle P_i \rangle$. Furthermore, the variation of $P^{(i)}$ from $\langle P^{(i)} \rangle$ decreases with increasing $\Delta l_\alpha$. Figures \ref{fig:IDAA}(g1)-\ref{fig:IDAA}(g3) show plots of the individual spiking measure $M_s^{(i)}$ versus the in-degree $d^{(in)}$ for $\Delta l_\alpha=$-15, 0, and 15, respectively. The value of individual spiking measure $M_s^{(i)}$ is determined by competition between the firing degree $F^{(i)}$ and the pacing degree $P^{(i)}$ of individual neurons, because $M_s^{(i)}$ is given by the product of both $F^{(i)}$ and $P^{(i)}$ [see Eq.~(\ref{eq:SM3})]. For more clear comparison and presentation, we normalize the firing degree $F^{(i)}$, the pacing degree $P^{(i)}$, and the spiking measure $M_s^{(i)}$ by dividing them with their ensemble-averaged values for the symmetric case of $\Delta l_\alpha=0$. The normalized firing degree ${\widetilde{F}}^{(i)}$, pacing degree ${\widetilde{P}}^{(i)}$, and spiking measure $\widetilde{M}_s^{(i)}$ are shown in Figs.~\ref{fig:IDAA}(h1)-\ref{fig:IDAA}(h3), Figs.~\ref{fig:IDAA}(i1)-\ref{fig:IDAA}(i3), and Figs.~\ref{fig:IDAA}(j1)-\ref{fig:IDAA}(j3). When compared with the case of $\Delta l_\alpha=0$, for $\Delta l_\alpha = -15$ the normalized ensemble-averaged firing degree $\langle {\widetilde{F}}^{(i)} \rangle$ increases, while the normalized ensemble-averaged pacing degree $\langle {\widetilde{P}}^{(i)} \rangle$ decreases a little more. Consequently, the normalized ensemble-averaged spiking measure $\langle \widetilde{M}_s^{(i)} \rangle$ becomes less than that for $\Delta l_\alpha=0$. On the other hand, for $\Delta l_\alpha = 15$ $\langle {\widetilde{F}}^{(i)} \rangle$ decreases, while $\langle {\widetilde{P}}^{(i)} \rangle$ increases only a little. As a result, $\langle \widetilde{M}_s^{(i)} \rangle$ also becomes less than that for $\Delta l_\alpha=0$.
However, it is a little greater than that for $\Delta l_\alpha=-15$ because the variation from the symmetric case of $\Delta l_\alpha=0$ is smaller for the case of $\Delta l_\alpha=15$. This normalized ensemble-averaged spiking measure $\langle \widetilde{M}_s^{(i)} \rangle$ of individual neurons corresponds to the normalized population spiking measure $\widetilde{M}_s$ shown in Fig.~\ref{fig:AA}(g). Based on the individual dynamics, it is found that the population spiking measure $M_s$ has its peak value for the case of symmetric attachment due to perfect matching between the inward and the outward edges. As the magnitude $| \Delta l_\alpha |$ of the asymmetry parameter is increased from 0, $M_s$ decreases in both directions because of mismatching between the inward and the outward edges. However, for the cases of both signs ($+/-$) with the same magnitude (e.g., $\Delta l_\alpha=15$ and -15) the values of $M_s$ are different, although their network topology such as $L_p$ and $C_b$ are the same. As shown above, $M_s$ for the case of positive asymmetric attachment with $\Delta l_\alpha=15$ is larger than that for the case of negative asymmetric attachment with $\Delta l_\alpha=-15$ due to the difference in the distributions of the in-degrees.

As the third case of network architecture, we consider the $\beta$-process (occurring with the probability $\beta$), in addition to the above $\alpha$-process (which occurs with the probability $\alpha$) ($\alpha + \beta =1$). Unlike the case of $\alpha$-process, no new nodes are added, and symmetric preferential attachments with the same in- and out-degrees [$l_{\beta}^{(in)} = l_{\beta}^{(out)} (\equiv l_{\beta}$)] are made between $l_{\beta}$ pairs of (pre-existing) source and target nodes which are also preferentially chosen according to the attachment probabilities $\Pi_{source}(d_i^{(out)})$ and $\Pi_{target}(d_i^{(in)})$ of Eq.~(\ref{eq:AP}), respectively, such that self-connections (i.e., loops) and duplicate connections (i.e., multiple edges) are excluded, as shown in  Fig.~1(b). Here we set $l_\beta=5$. We investigate the effect of the $\beta$-process on sparse synchronization by varying $\beta$ for the three cases of $\Delta l_\alpha=$-15, 0, and 15. Figures \ref{fig:BP}(a1)-\ref{fig:BP}(a3) show the raster plots of spikes for $\beta=$0, 0.6, and 0.8, respectively in the case of $\Delta l_\alpha=-15$. As $\beta$ is increased from 0, the stripes in the raster plot become more clear, and the IPSR kernel estimates $R(t)$ show larger-amplitude regular oscillations, as shown in Figs.~\ref{fig:BP}(b1)-\ref{fig:BP}(b3). Also for both cases of $\Delta l_\alpha=0$ and 15, similar effect of $\beta$-process occurs in the raster plots of spikes and the IPSR kernel estimates $R(t)$, as shown in Figs.~\ref{fig:BP}(c1)-\ref{fig:BP}(f3). Consequently, with increasing $\beta$ the degree of sparse synchronization becomes better. For characterization of the effect of $\beta$ on the network topology, we also measure the average path length $L_p$ and the betweenness centralization $C_b$ by varying $\beta$. Figures \ref{fig:BP}(g) and \ref{fig:BP}(h) show the plots of $L_p$ and $C_b$ versus $\beta$, respectively for the three cases of $\Delta l_\alpha=-15$, 0, and 15. As $\beta$ is increased, both $L_p$ and $C_b$ decrease monotonically for all three cases of $\Delta l_\alpha$. As explained above, decrease in $L_p$ leads to reduction in total centrality $B_{tot}$ (i.e., the sum of centralities of all nodes). How $B_{tot}$ decreases with increase in $\beta$ can be seen explicitly in Figs.~\ref{fig:BP}(i1)-\ref{fig:BP}(i3) for the cases of $\Delta l_\alpha=$-15, 0, and 15, respectively. We note that the maximum betweenness $B_{max}$ of the head hub is much more reduced than the average centralities of the secondary hubs and the peripheral nodes, ${\langle B \rangle}_{hub}$ and ${\langle B \rangle}_{peri}$, for each case of $\Delta l_\alpha$, which results in decrease in differences between $B_{max}$ of the head hub and $B_i$ of other nodes (i.e., decrease in $C_b$). Hence, with increasing $\beta$, typical separation between two nodes in the network becomes shorter and load of communication traffic becomes less concentrated on the head hub. Thus, as $\beta$ is increased, efficiency of communication between nodes becomes better, which may result in increase in the degree of sparse synchronization. Figures \ref{fig:BP}(j1)-\ref{fig:BP}(j2) show plots of the average occupation degree $\langle O_i \rangle$ and the average pacing degree $\langle P_i \rangle$ versus $\beta$. As $\beta$ is increased, at first $\langle O_i \rangle$ decreases for both cases of $\Delta l_\alpha=$-15 and 0, while it increases very little for $\Delta l_\alpha=15$. Then, they seem to approach each other for large $\beta$. On the other hand, $\langle P_i \rangle$ increases markedly for all the three cases of $\Delta l_\alpha$. Consequently, the statistical-mechanical spiking measure $M_s$, given by taking into consideration both the occupation and the pacing degrees, increases monotonically mainly due to marked increase in $\langle P_i \rangle$ for all three cases of $\Delta l_\alpha$, as shown in Fig.~\ref{fig:BP}(j3). 

Finally, we investigate contributions of individual neuronal dynamics to sparse synchronization by varying $\beta$ for the three cases of $\Delta l_\alpha=$-15, 0, and 15. Figures \ref{fig:IDBP}(a1)-\ref{fig:IDBP}(a3) show ``comet-shaped'' plots of the out-degree $d^{(out)}$ versus the in-degree $d^{(in)}$ for $\beta=$0, 0.6, and 0.8 in the case of $\Delta l_\alpha$= -15. For each $\beta$, peripheral nodes (correspond to the coma part of the comet) are enclosed by the rectangle, while hubs (corresponding to the tail part of the comet) lie outside the rectangle and the head hub (node 1) with the highest degree is represented by the open circle. In the $\beta-$ process, the probability that the head hub may be chosen as a source and/or a target node is low because self-connections (i.e., loops) and duplicate connections are excluded. Hence, there is no particular change in the degree of the head hub, unlike the case of $\alpha-$process in Fig.~\ref{fig:ESFN}. On the other hand, there is a marked increase in the degrees of some (pre-existing) peripheral nodes and secondary hubs through the $\beta-$process, which results in the immigration of some peripheral nodes into the secondary hub group. As a result, with increasing $\beta$ the tail part of the comet is intensified (i.e., the secondary hub group is intensified) because the number of secondary hubs is increased. Then, although the number of peripheral nodes is reduced, the size of the coma part (i.e., the size of the rectangle enclosing the peripheral group) increases with increasing $\beta$ because both the in- and the out-degrees of peripheral nodes are increased. Figures \ref{fig:IDBP}(d1)-\ref{fig:IDBP}(d3) also show the power-law distributions of in-degree $d^{(in)}$ for $\beta=$0, 0.6, and 0.8, respectively. As $\beta$ is increased, the exponent $\gamma$ decreases, because the secondary hub group is intensified (i.e. their fraction of nodes is increased) but the fraction of peripheral nodes is decreased. For the other two cases of $\Delta l_\alpha=0$ and 15
(see the middle and the right panels in Fig.~\ref{fig:IDBP}), as $\beta$ is increased the distributions of in- and out-degrees evolve in a similar way, as shown in Figs.~\ref{fig:IDBP}(b1)-\ref{fig:IDBP}(b3) and Figs.~\ref{fig:IDBP}(c1)-\ref{fig:IDBP}(c3), respectively. The main effect of $\beta-$process is to intensify the secondary hub group (without particular change in the degree of the head hub) (i.e., the tail part of the comet-shaped distribution is intensified with increasing $\beta$). Hence, as $\beta$ is increased the exponent for the power-law distributions of in-degree decreases [see Figs.~\ref{fig:IDBP}(e1)-\ref{fig:IDBP}(e3) and Figs.~\ref{fig:IDBP}(f1)-\ref{fig:IDBP}(f3)]. These in-degree distributions for $\Delta l_\alpha=-15$, 0, and 15 affect the MFRs of individual neurons. Based on the change in the in-degree distribution in the $\beta-$process, we study the effect of the $\beta-$process on the MFRs of individual neurons. Figures \ref{fig:IDBP}(g1)-\ref{fig:IDBP}(g3) show the distribution of MFRs of individual neurons versus the in-degree $d^{(in)}$ for $\beta=$0, 0.6, and 0.8 in the case of $\Delta l_\alpha$= -15. For each $\beta$, the  ensemble-averaged MFR is represented by the horizontal gray line. As $\beta$ is increased, the ensemble-averaged in-degree of the peripheral neurons increases, and hence the size of the subset of pre-synaptic neurons of a typical peripheral neuron becomes larger. Then, the variation of MFRs of peripheral neurons becomes reduced particularly because the part of higher MFRs gradually disappears. Consequently, the ensemble-averaged MFR in the whole population decreases a little due to the majority of peripheral neurons. With increasing $\beta$ the number of hubs with higher in-degrees increases, and each hub receives less inhibition on average because of the decreased ensemble-averaged MFR. Consequently, distribution of MFRs of the hubs goes upward and approaches the ensemble-averaged gray line. For the symmetric case of $\Delta l_\alpha=0$, distribution of MFRs of individual neurons also evolves in a similar way with increasing $\beta$, as shown in Figs.~\ref{fig:IDBP}(h1)-\ref{fig:IDBP}(h3). For $\beta=0$ the distribution of MFRs of peripheral neurons becomes more narrow than that for $\Delta l_\alpha=-15$ [compare Figs.~\ref{fig:IDBP}(h1) with \ref{fig:IDBP}(g1)], due to increased average in-degree of peripheral neurons (refer to a detailed explanation in Fig.~\ref{fig:IDAA}). As $\beta$ is increased, the average in-degree of peripheral neurons also increases more, and hence the variation of MFRs of peripheral neurons becomes more reduced. Consequently, the ensemble-averaged MFR in the whole population decreases very little with increasing $\beta$, in comparison to the case of $\Delta l_\alpha=-15$. As $\beta$ is increased, the distribution of MFRs of hubs goes more upward than those for $\Delta l_\alpha=-15$ because the hubs receive less synaptic inhibition. For the case of positive asymmetric attachment with $\Delta l_\alpha=15$, evolution of distribution of MFRs of individual neurons also follows a similar way with increasing $\beta$, as shown in Figs.~\ref{fig:IDBP}(i1)-\ref{fig:IDBP}(i3). For $\beta=0$, distribution of MFRs for $\Delta l_\alpha=15$ is much narrower than those for $\Delta l_\alpha=-15$ and 0, and hence with increase in $\beta$ the distribution of MFRs of peripheral neurons becomes reduced a little more. Then, the average MFR of peripheral neurons decreases a little. On the other hand, the number of hubs increases, their distribution of MFRs goes upward, and hence the average MFR of hubs increases. For this case, the effect of hubs is a little greater than that of peripheral neurons, and hence the ensemble-averaged MFR in the whole population becomes increased very little. Based on these distributions of MFRs of individual neurons, we also study contributions of individual dynamics to the sparse synchronization by varying $\beta$. Figures \ref{fig:IDBP}(j1)-\ref{fig:IDBP}(j3) show plots of the firing degree $F^{(i)}$ of individual neurons versus the in-degree $d^{(in)}$ for $\beta=$0, 0.6, and 0.8 in the case of $\Delta l_\alpha=-15$, respectively. As mentioned in Fig.~\ref{fig:IDAA}, distribution of the individual firing degree $F^{(i)}$ is strongly correlated with their distribution of MFRs. As in the case of MFRs, $F^{(i)}$ also spreads around the ensemble-averaged value $\langle F^{(i)} \rangle$ [denoted by the gray line and corresponding to the average occupation degree $\langle O_i \rangle$ (denoted by the triangles) in Fig.~\ref{fig:BP}(j1)] which decreases with increasing $\beta$. For both cases of increased $\Delta l_\alpha=0$ and 15, distributions of $F^{(i)}$ also evolve in a similar way with increasing $\beta$, as shown in Figs.~\ref{fig:IDBP}(k1)-\ref{fig:IDBP}(k3) and Figs.~\ref{fig:IDBP}(l1)-\ref{fig:IDBP}(l3), respectively. For $\Delta l_\alpha=0$, $\langle F^{(i)} \rangle$ decreases a little as $\beta$ is increased, while  $\langle F^{(i)} \rangle$ for $\Delta l_\alpha=15$ increases a little. These results of $\langle F^{(i)} \rangle$ lead to variation of $\langle O_i \rangle$ in Fig.~\ref{fig:BP}(j1). Distributions of the pacing degree $P^{(i)}$ of individual neurons for $\Delta l_\alpha=-15$, 0, and 15 also show spreads from their ensemble-averaged values (represented by gray lines), as shown in Figs.~\ref{fig:IDBP}(m1)-\ref{fig:IDBP}(m3), Figs.~\ref{fig:IDBP}(n1)-\ref{fig:IDBP}(n3), and Figs.~\ref{fig:IDBP}(o1)-\ref{fig:IDBP}(o3), respectively. For each $\Delta l_\alpha$, the variance in the distribution of MFRs decreases with increasing $\beta$, and hence the ensemble-averaged pacing degree $\langle P^{(i)} \rangle$, corresponding to the average pacing degree $\langle P_i \rangle$ in Fig.~\ref{fig:BP}(j2), increases monotonically as $\beta$ is increased, while the variation of $P^{(i)}$ from $\langle P^{(i)} \rangle$ tends to decrease. Figures \ref{fig:IDBP}(p1)-\ref{fig:IDBP}(p3), Figs.~\ref{fig:IDBP}(q1)-\ref{fig:IDBP}(q3), and Figs.~\ref{fig:IDBP}(r1)-\ref{fig:IDBP}(r3) show plots of the individual spiking measure $M_s^{(i)}$ versus the in-degree $d^{(in)}$ for $\Delta l_\alpha=$-15, 0, and 15, respectively. For all three cases of $\Delta l_\alpha$, with increasing $\beta$ $\langle P^{(i)} \rangle$ increases markedly, in comparison to $\langle F^{(i)} \rangle$. Consequently, for each $\Delta l_\alpha$ the ensemble-averaged spiking measure $\langle M_s^{(i)} \rangle$, corresponding to the population spiking measure $M_s$ in Fig.~\ref{fig:BP}(j3), increases as $\beta$ is increased.

In the way explained above, for all the three cases of effect of network architecture on sparse synchronization, contributions of individual neuronal dynamics on population synchronization vary depending on the in-degrees, although their ensemble-averaged individual measures, $\langle F^{(i)} \rangle$, $\langle P^{(i)} \rangle$, and $\langle M_s^{(i)} \rangle$, give the population measures such as $\langle O_i \rangle$, $\langle P_i \rangle$, and $M_s$. Consequently, dynamics of individual neurons for the case of sparse synchronization reveal the inhomogeneous structure of the SFN, in contrast to statistically homogeneous random and small-world networks.

\section{Summary}
\label{sec:SUM}
In order to extend previous works on sparse synchronization in statistically homogeneous networks such as random graphs and small-world networks \cite{Sparse1,Sparse2,Sparse3,Sparse4,Sparse5,Sparse6,Kim} to the case of inhomogeneous networks, we have investigated emergence of sparsely synchronized brain rhythms in the directed version of the Barab\'{a}si-Albert SFN model with symmetric preferential attachment with the same in- and out-degrees $(l_{\alpha}^{(in)} = l_{\alpha}^{(out)} \equiv l_{\alpha}=25)$. Fast sparsely synchronized rhythms with stochastic and intermittent neuronal discharges have been found to emerge for large values of $J$ and $D$. We have made an intensive investigation of population states by varying $D$ for a fixed value of $J =1500$. For small $D$, fully synchronized rhythms with the same $f_p$ (population-rhythm frequency) and $f_i$ (MFR of individual neurons) appear. For this case of full synchronization, all the individual neurons exhibit the same oscillatory behaviors, independently of inhomogeneous network structure. However, as $D$ passes a lower threshold $D_{th,l} (\simeq 109)$, partial synchronization with $f_p > {\langle f_i \rangle}$ (ensemble-averaged MFR of individual neurons) occurs due to intermittent discharge of individual neurons. As $D$ is increased from $D_{th,l}$, difference between $f_p$ and ${\langle f_i \rangle}$ increases, and when passing a higher threshold $D_{th,h} (\simeq 400)$, sparsely synchronized rhythms with $f_p > 4 {\langle f_i \rangle}$ appear. Unlike the case of full synchronization, MFRs of individual neurons vary depending on their in-degrees for the case of partial and sparse synchronization. As $D$ is further increased and eventually passes a critical value $D^*(\simeq 759)$, a transition to unsynchronization occurs due to destructive role of noise to spoil the pacing between sparse spikes. The critical value $D^*$ has been determined through calculation of the thermodynamic order parameter $\cal{O}$. For $D < D^*$, population synchronization has been found to emerge because the spatial correlation length between the neuronal pairs covers the whole system. Moreover, the degree of population synchronization has also been measured in terms of two types of statistical-mechanical spiking and correlation measures. Unlike the case of full synchronization, individual neuronal dynamics vary depending on their in-degrees and reveal the inhomogeneous network structure for the case of partial and sparse synchronization, in contrast to the case of statistically homogeneous random graphs and small-world networks. As a next step, we have also investigated the effect of network architecture on sparse synchronization for fixed values of $J =1500$ and $D =450$ in the following three cases: (1) variation in the degree of symmetric attachment (2) asymmetric preferential attachment of new nodes with different in- and out-degrees (3) preferential attachment between pre-existing nodes (without addition of new nodes). As the degree $l_\alpha$ of symmetric preferential attachment is increased, both the average path length $L_p$ and the betweenness centralization $C_b$ have been found to decrease. Hence, typical separation between two nodes in the network becomes shorter and load of communication traffic becomes less concentrated on the head hub. Consequently, with increasing $l_\alpha$ the degree of sparse synchronization has been found to become higher due to increased efficiency of communication between nodes. On the other hand, the normalized axon wire length ${\cal{L}}_w$ of the network also increases. Through a trade-off between the population synchronization and the wiring economy, an optimal sparsely-synchronized rhythm has been found to emerge at a minimal wiring cost in an economic SFN with an optimal degree $l_\alpha^* (\simeq 34)$. As the second case of network architecture, we have also considered the case of asymmetric preferential attachment of new nodes with different in- and out-degrees (i.e., $l_{\alpha}^{(in)} \neq l_{\alpha}^{(out)}$). For this case, we have also measured $L_p$ and $C_b$ by varying the asymmetry parameter $\Delta l_\alpha$ denoting the deviation from the above symmetric case, and investigated how the sparse synchronization changes. As the magnitude $| \Delta l_\alpha |$ of the asymmetry parameter increases, both $L_p$ and $C_b$ have been found to increase symmetrically, independently of the sign of $\Delta l_\alpha$. Hence, with increasing $| \Delta l_\alpha |$ typical separation between two nodes in the network becomes longer and load of communication traffic becomes more concentrated on the head hub, due to increased mismatching between the inward and outward edges. Consequently, as $| \Delta l_\alpha |$ is increased the degree of sparse synchronization has been found to become lower due to decreased efficiency of communication between nodes. For both cases of the positive and the negative asymmetries with the same magnitude (e.g., $\Delta l_\alpha = 15$ and -15), the values of $L_p$ and $C_b$ have been found to be nearly the same, because both inward and outward edges are involved equally in computation of $L_p$ and $C_b$. However, their degrees of sparse synchronization have been found to become different due to their distinctly different in-degree distributions affecting individual MFRs. In addition to the above $\alpha$-process where preferential attachment is made to newly added nodes with probability $\alpha$, we have also considered another $\beta$-process where preferential attachment between pre-existing nodes (without addition of new nodes) is made with probability $\beta$ ($\alpha + \beta =1$). By varying the probability $\beta$, we have also measured $L_p$ and $C_b$ and investigated the effect of this $\beta$-process on sparse synchronization. As $\beta$ is increased, communication between pre-existing neurons becomes more efficient due to decrease in both $L_p$ and $C_b$, and consequently the degree of sparse synchronization has been found to increase. For these three cases of network architecture, contributions of individual neuronal dynamics on the sparse synchronization were also characterized in terms of their MFRs, firing degrees, pacing degrees, and spiking measures. It has thus been found that the dynamics of individual neurons vary depending on their in-degrees and reveal the inhomogeneous structure of the SFN, in contrast to the statistically homogeneous networks such as the random graph and the small-world network. Finally, we expect that our results might provide important insights on emergence of fast sparsely synchronized rhythms, associated with diverse cognitive functions such as sensory perception, feature integration, selective attention, and memory formation, in real brain networks with scale-free property.

\begin{acknowledgments}
This research was supported by Basic Science Research Program through the National Research Foundation of Korea(NRF) funded by the Ministry of Education (Grant No. 2013057789).
\end{acknowledgments}

\newpage
\begin{figure}
\includegraphics[width=\columnwidth]{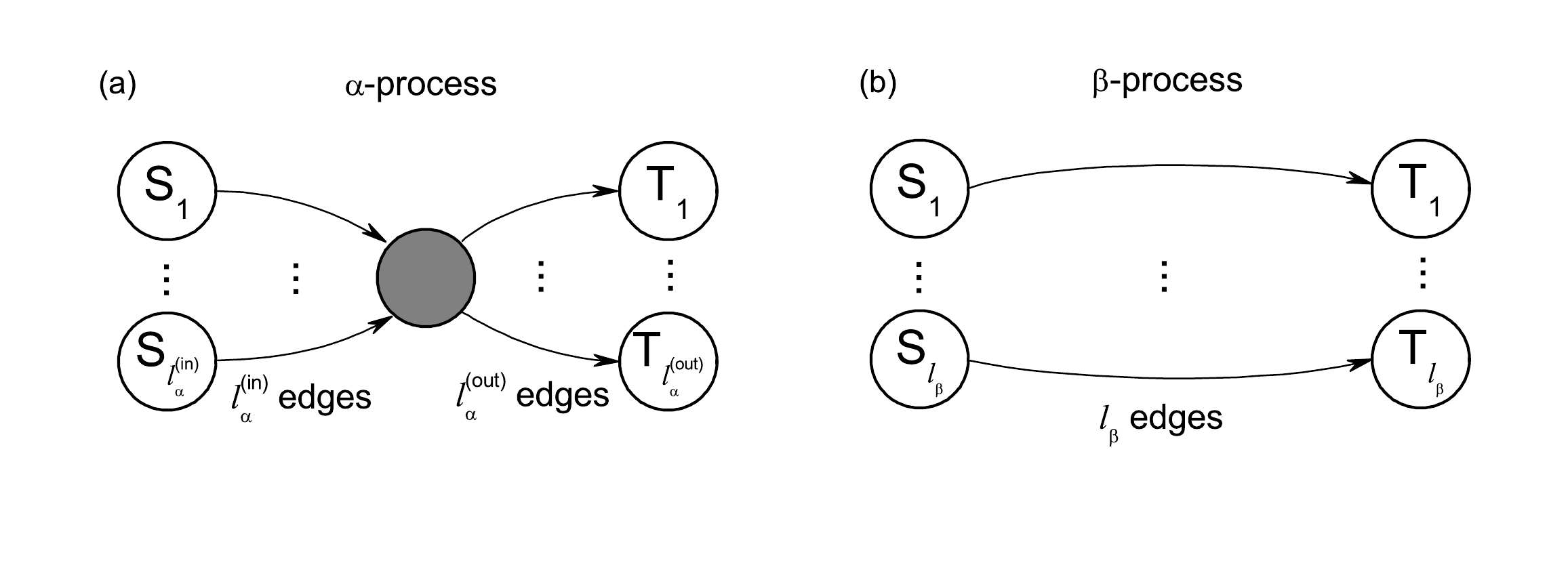}
\caption{Diagrams of two processes generating a directed SFN. (a) Diagram of the $\alpha$-process of adding a new node (denoted by a gray circle) with preferential attachment of $l_{\alpha}^{(in)}$ inward and $l_{\alpha}^{(out)}$ outward edges. (b) Diagram of the $\beta-$process of the preferential attachment between $l_{\beta}$ pairs of (pre-existing) source and target nodes without adding a new node. The open circles with labels ``$S$'' and ``$T$'' represent the (pre-existing) source and target nodes, respectively.
}
\label{fig:SFN}
\end{figure}

\newpage
\begin{figure}
\includegraphics[width=\columnwidth]{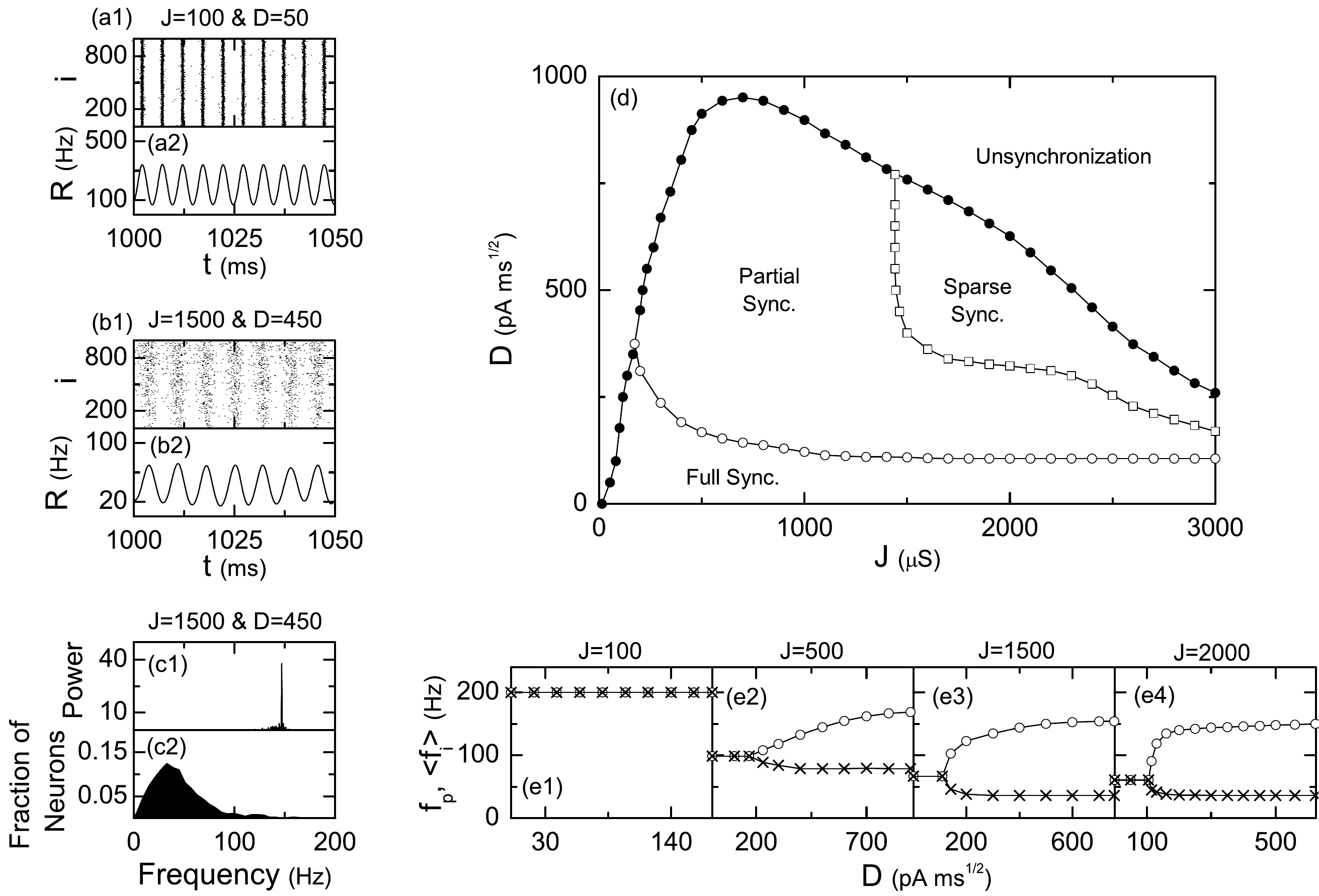}
\caption{State diagram in the $J-D$ plane for $I_{DC}=1500$ in the directed SFN of $\alpha =1$ (i.e., $\beta=0$) and  $l_{\alpha}^{(in)} = l_{\alpha}^{(out)} \equiv l_{\alpha} =25$ (i.e., symmetric preferential attachment).
(a1) Raster plot of spikes and (a2) plot of the IPSR kernel estimate $R(t)$ for the full synchronization when $J=100$  and $D=50$. (b1) Raster plot of spikes and (b2) plot of the IPSR kernel estimate $R(t)$ for the sparse synchronization when $J=1500$ and $D=450$. The band width of the Gaussian kernel estimate for the IPSR $R(t)$ is 1 ms. (c1) One-sided power spectrum of $\Delta R(t)$ [$= R(t) - \overline{R(t)}]$  (the overbar represents the time average) with mean-squared amplitude normalization and (c2) distribution of mean firing rates (MFRs) of individual neurons for $J=1500$  and $D=450$. Power spectrum is obtained from $2^{16}$ (=65536) data points. Averaging time for the MFR is $10^4$ ms and the bin size for the histogram is 3 Hz. (d) State diagram in the $J-D$ plane. For the full synchronization, the individual MFR $f_i$ is the same as the population frequency $f_p$, while for the partial and sparse synchronization, the ensemble-averaged MFR $\langle f_i \rangle$ is less than $f_p$. Particularly, the case of $f_p > 4 {\langle f_i \rangle}$ is referred to as the sparse synchronization. Plots of $f_p$  and $\langle f_i \rangle$ versus for $J=$ (e1) 100, (e2) 500, (e3) 1500, and (e4) 2000. Here, the circles and the crosses denote $f_p$ and $\langle f_i \rangle$, respectively.
}
\label{fig:SD}
\end{figure}

\newpage
\begin{figure}
\includegraphics[width=\columnwidth]{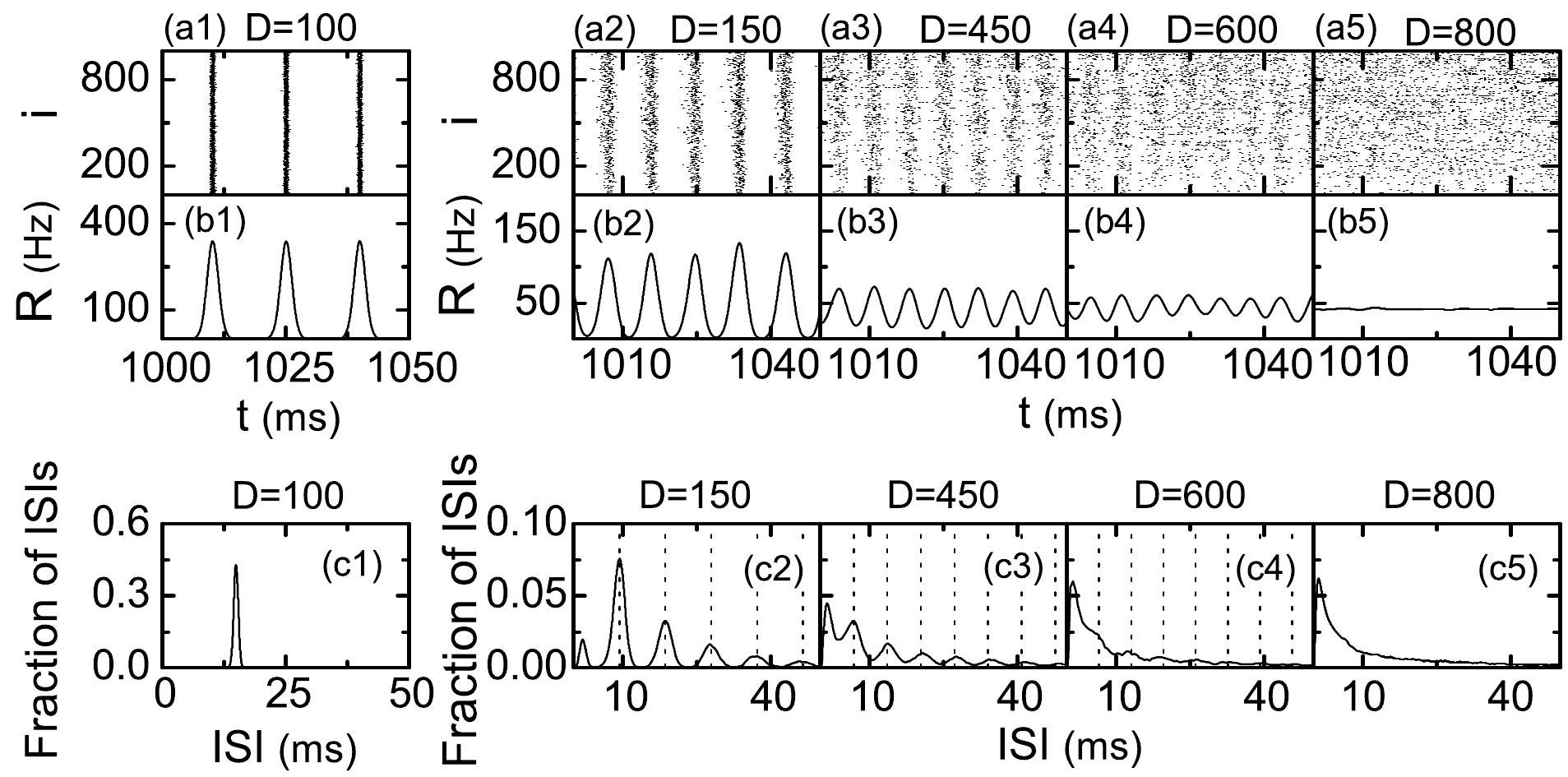}
\caption{Fast sparse synchronization for $I_{DC}=1500$ and $J=1500$ in the directed SFN of $\alpha =1$ (i.e., $\beta=0$) and  $l_{\alpha}^{(in)} = l_{\alpha}^{(out)} \equiv l_{\alpha} =25$ (i.e., symmetric preferential attachment). Raster plots of spikes in (a1)-(a5), plots of the IPSR kernel estimate $R(t)$ in (b1)-(b5), and the ISI histograms in (c1)-(c5) for various values of $D=$ 100, 150, 450, 600, and 800; vertical dotted lines denote integer multiples of the global period $T_G$ of $R(t)$ [$\simeq$ 9.3 ms in (c2), 6.8 ms in (c3), and 6.5 ms in (c4)]. The band width of the Gaussian kernel estimate for the IPSR $R(t)$ is 1 ms. Each ISI histogram is composed of
$5 \times 10^4$ ISIs and the bin size for the histogram is 0.5 ms.
}
\label{fig:FSS}
\end{figure}

\newpage
\begin{figure}
\includegraphics[width=0.8\columnwidth]{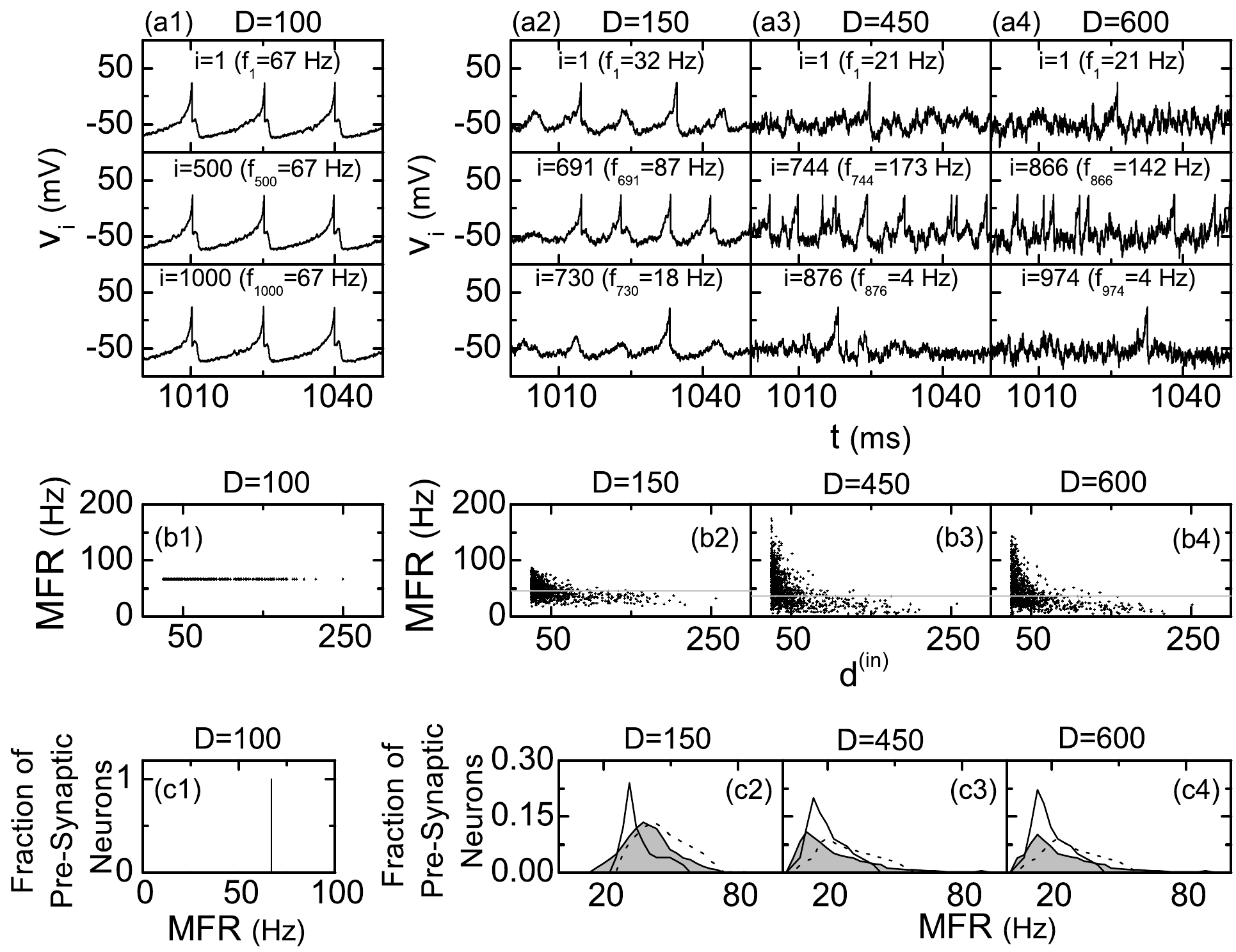}
\caption{Individual neuronal dynamics for $I_{DC}=1500$ and $J=1500$ in the directed SFN of $\alpha =1$ (i.e., $\beta=0$) and  $l_{\alpha}^{(in)} = l_{\alpha}^{(out)} \equiv l_{\alpha} =25$ (i.e., symmetric preferential attachment). (a1)-(a4) Time series of the membrane potentials of the hub neuron $(i=1)$ with highest degree and the fastest and the slowest peripheral neurons with low degrees for various values of $D=$ 100, 150, 450, and 600. (b1)-(b4) Plots of MFRs of individual neurons versus the in-degree $d^{(in)}$ for $D=$ 100, 150, 450, and 600. Averaging time for the MFR is $10^4$ ms. (c1)-(c4) Histograms for the MFRs of pre-synaptic neurons for the hub neuron and the fastest and the slowest peripheral neurons with low degrees for $D=$ 100, 150, 450, and 600. Horizontal gray lines in (b2)-(b4) denote ensemble-averaged MFRs [$\simeq$ 46 Hz in (b2) and 36 Hz in (b3)-(b4)]. Gray regions, solid lines, and dotted lines in (c2)-(c4) denote the histograms for the MFRs of the pre-synaptic neurons for the hub neuron and the fastest and the slowest peripheral neurons with low degrees, respectively. Histograms in (c1)-(c4) are obtained from 30 realizations and the bin size for the histogram is 3 Hz.
}
\label{fig:ID}
\end{figure}

\newpage
\begin{figure}
\includegraphics[width=\columnwidth]{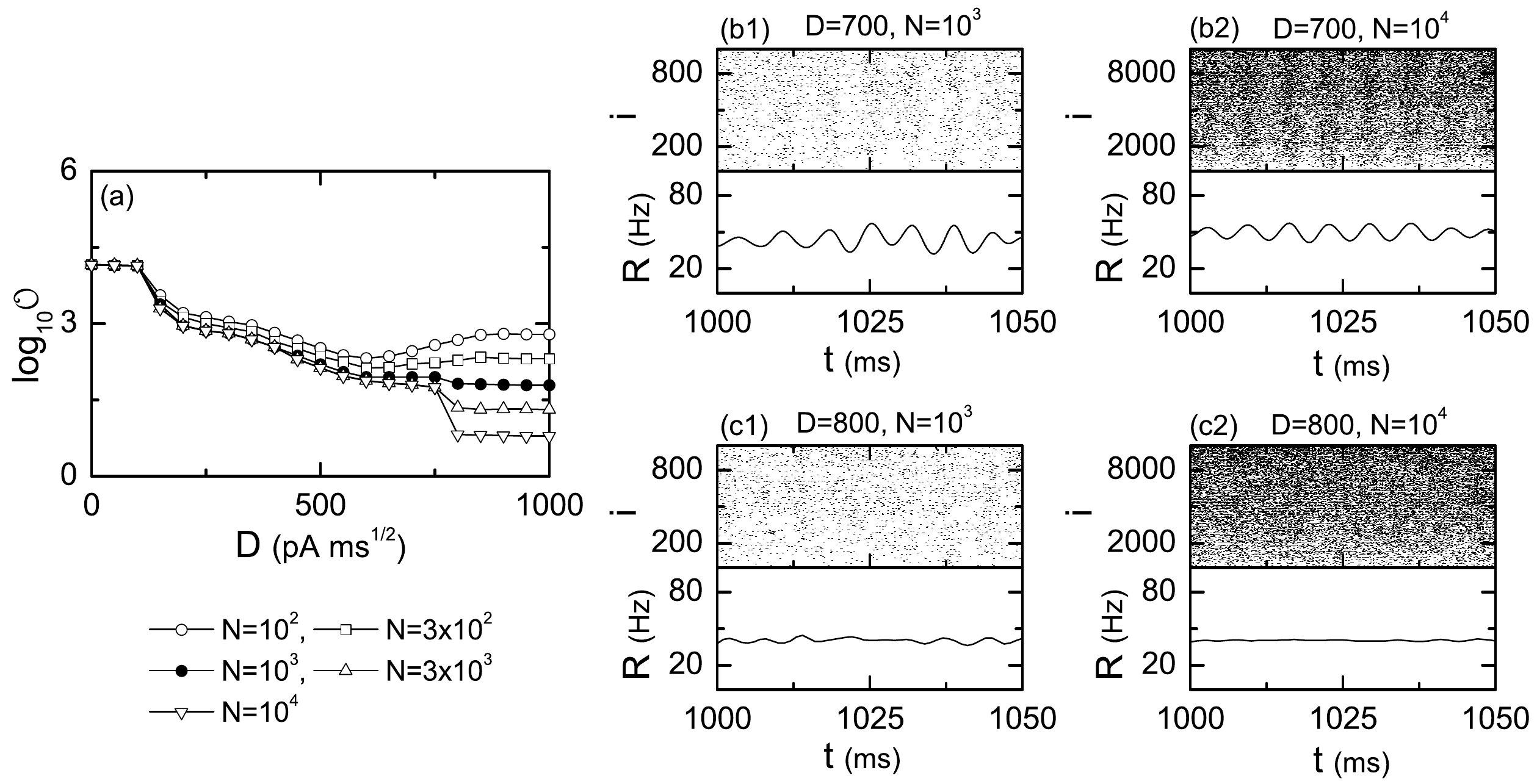}
\caption{Synchronization-unsynchronization transition for $I_{DC}=1500$ and $J=1500$ in the directed SFN of $\alpha =1$ (i.e., $\beta=0$) and  $l_{\alpha}^{(in)} = l_{\alpha}^{(out)} \equiv l_{\alpha} =25$ (i.e., symmetric preferential attachment). (a) Plots of the thermodynamic order parameter versus $D$. Averaging time for the thermodynamic order parameter is $3 \times 10^4$ ms. Synchronized state for $D=700$: raster plots of spikes and plots of the IPSR kernel estimate $R(t)$ for $N=$ (b1) $10^3$  and (b2) $10^4$. Unsynchronized state for $D=800$: raster plots of spikes and plots of the IPSR kernel estimate $R(t)$ for $N=$ (c1) $10^3$ and (c2) $10^4$. The band width of the Gaussian kernel estimate for the IPSR is 1 ms.
}
\label{fig:Order}
\end{figure}

\newpage
\begin{figure}
\includegraphics[width=0.8\columnwidth]{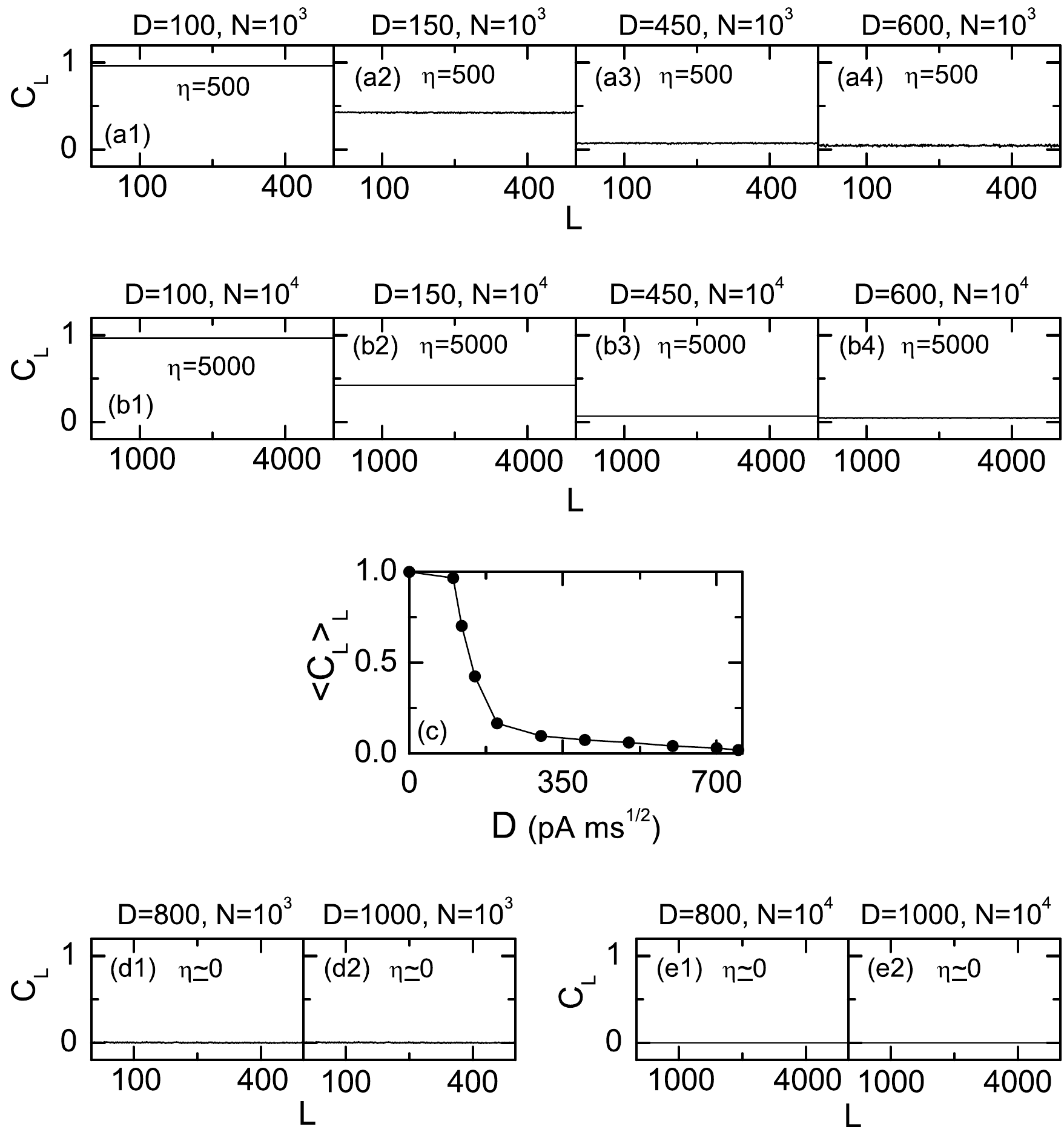}
\caption{Characterization of synchronization-unsynchronization transition in terms of spatial cross-correlations for $I_{DC}=1500$ and $J=1500$ in the directed SFN of $\alpha =1$ (i.e., $\beta=0$) and  $l_{\alpha}^{(in)} = l_{\alpha}^{(out)} \equiv l_{\alpha} =25$ (i.e., symmetric preferential attachment). Plots of the spatial correlation function $C_L$ between neuronal pairs versus spatial distance $L$ for the synchronized cases of various values of $D=$100, 150, 450, and 600 when $N=$ (a1)-(a4) $10^3$  and (b1)-(b4) $10^4$. (c) Plot of the average spatial-correlation degree ${\langle C_L \rangle}_L$ versus $D$. Plots of the spatial correlation function $C_L$ versus $L$ for the unsynchronized cases of $D=800$ and 1000 when (d1)-(d2) $N=10^3$ and (e1)-(e2) $N=10^4$. The number of data used for the calculation of each temporal cross-correlation function $C_{i,j}(\tau)$ (the values at zero-time lag ($\tau=0$) are used for calculation of $C_L$) is $2 \times 10^4$.
}
\label{fig:SC}
\end{figure}

\newpage
\begin{figure}
\includegraphics[width=\columnwidth]{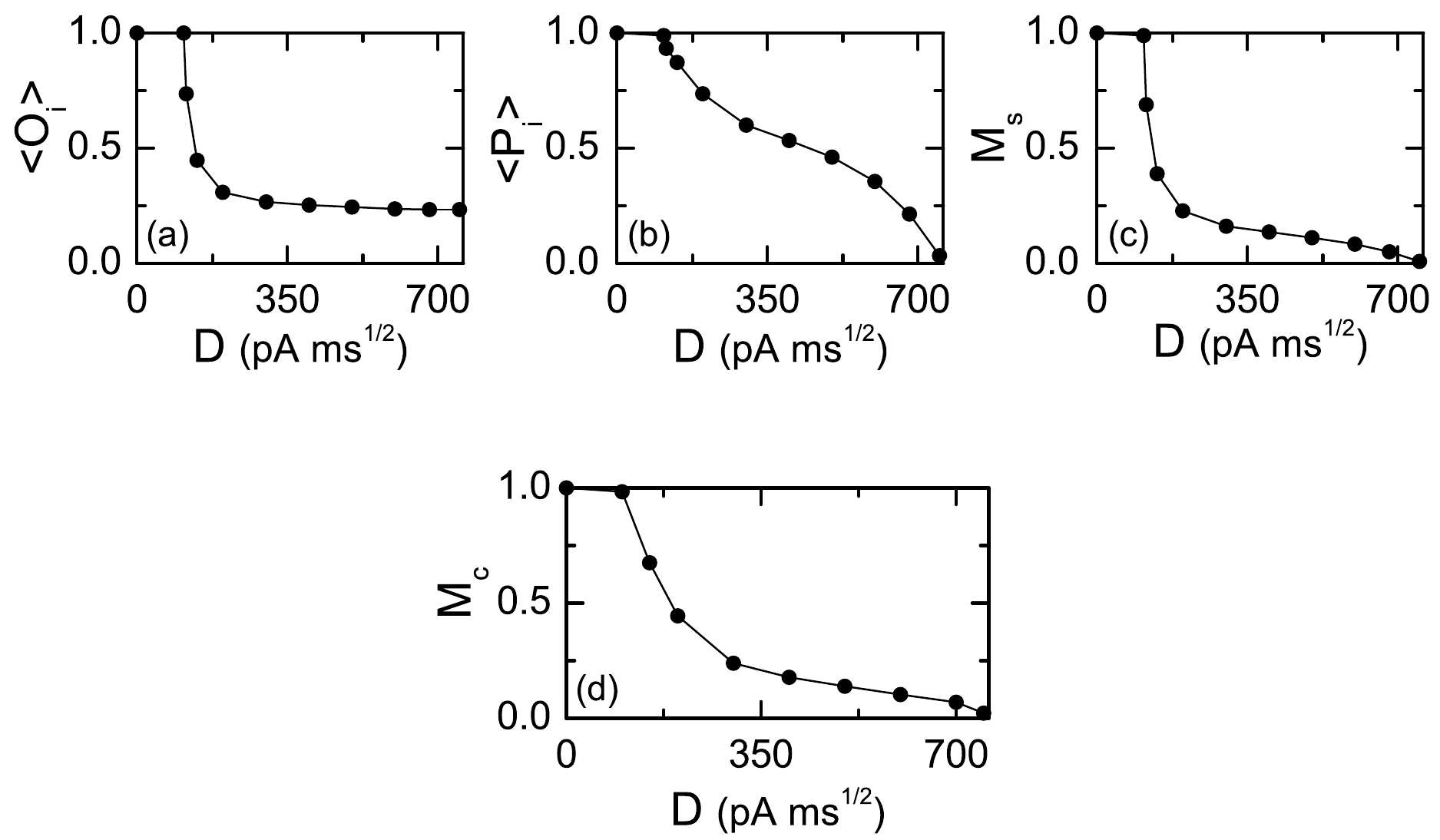}
\caption{Characterization of population synchronization for $I_{DC}=1500$ and $J=1500$ in the directed SFN of $\alpha =1$ (i.e., $\beta=0$) and  $l_{\alpha}^{(in)} = l_{\alpha}^{(out)} \equiv l_{\alpha} =25$ (i.e., symmetric preferential attachment). Plots of (a) the average occupation degree $\langle O_i \rangle$, (b) the average pacing degree $\langle P_i \rangle$, and (c) the statistical-mechanical spiking measure $M_s$ versus $D$. $\langle O_i \rangle$, $\langle P_i \rangle$, and $M_s$ are obtained by following $3 \times 10^3$ stripes in the raster plot of spikes. (d) Plot of the statistical-mechanical correlation measure $M_c$, based on temporal cross-correlations between the IPSR $R(t)$ and IISRs $r_i(t)$ of individual neurons versus $D$. The number of data used for the calculation of temporal cross-correlation function for each $D$ is $2 \times 10^4$.
}
\label{fig:SM}
\end{figure}

\newpage
\begin{figure}
\includegraphics[width=\columnwidth]{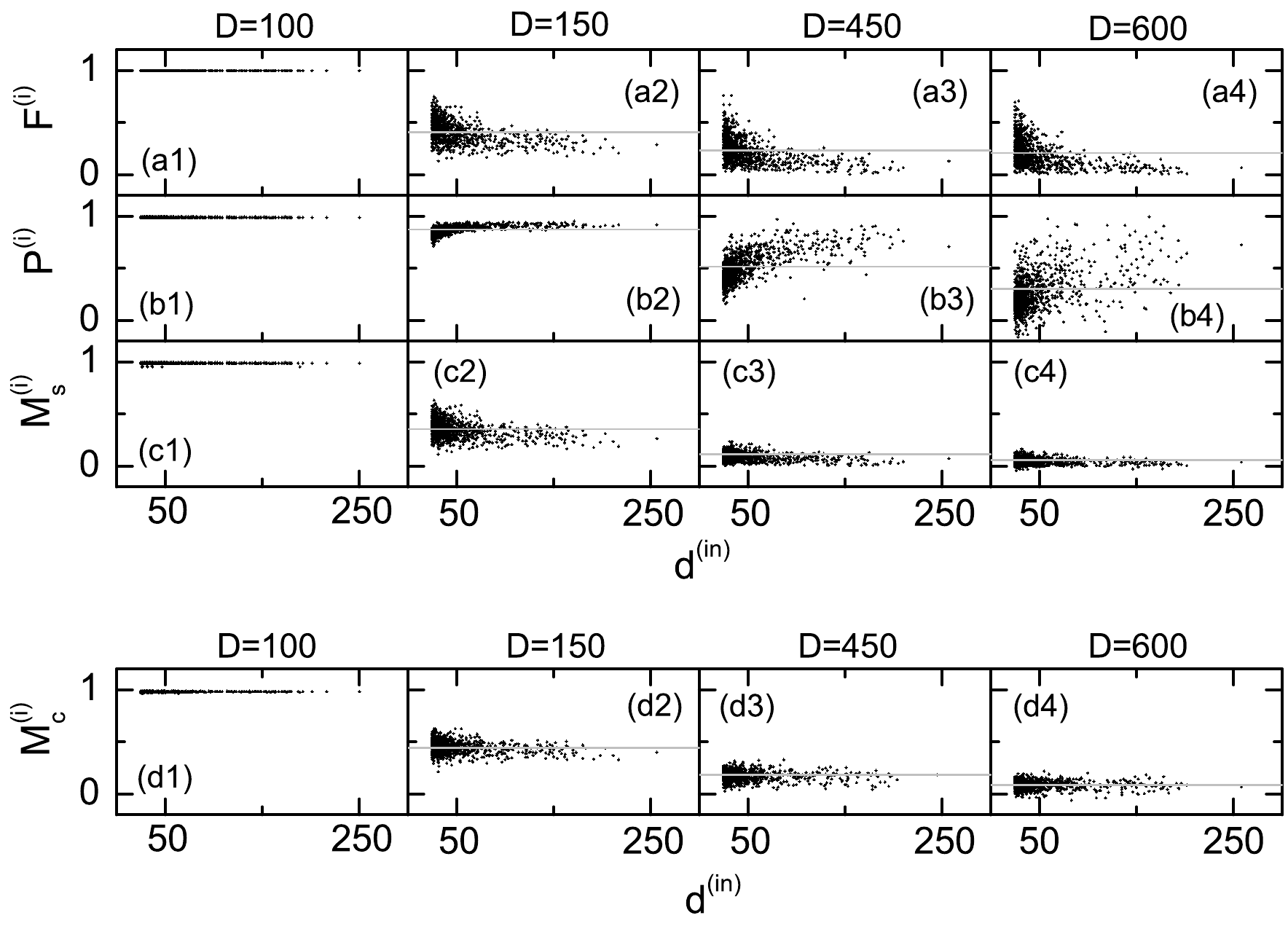}
\caption{Contributions of individual dynamics to the population synchronization for $I_{DC}=1500$ and $J=1500$ in the directed SFN of $\alpha =1$ (i.e., $\beta=0$) and  $l_{\alpha}^{(in)} = l_{\alpha}^{(out)} \equiv l_{\alpha} =25$ (i.e., symmetric preferential attachment). Plots of (a) the firing degree $F^{(i)}$, (b) the pacing degree $P^{(i)}$, and (c) the spiking measure $M_s^{(i)}$ of individual neurons versus the in-degree $d^{(in)}$ for various values of $D=$100, 150, 450, and 600. $F^{(i)}$, $P^{(i)}$, and $M_s^{(i)}$ are obtained by following $3 \times 10^3$ stripes in the raster plot of spikes. (d) Plots of cross-correlations between IPSR $R(t)$ and IISRs $r_i(t)$ of individual neurons versus the in-degree $d^{(in)}$ for $D=$100, 150, 450, and 600. The number of data used for the calculation of temporal cross-correlation function for each $D$ is $2 \times 10^4$. Horizontal gray lines represent ensemble-averaged values.
}
\label{fig:IM}
\end{figure}

\newpage
\begin{figure}
\includegraphics[width=0.65\columnwidth]{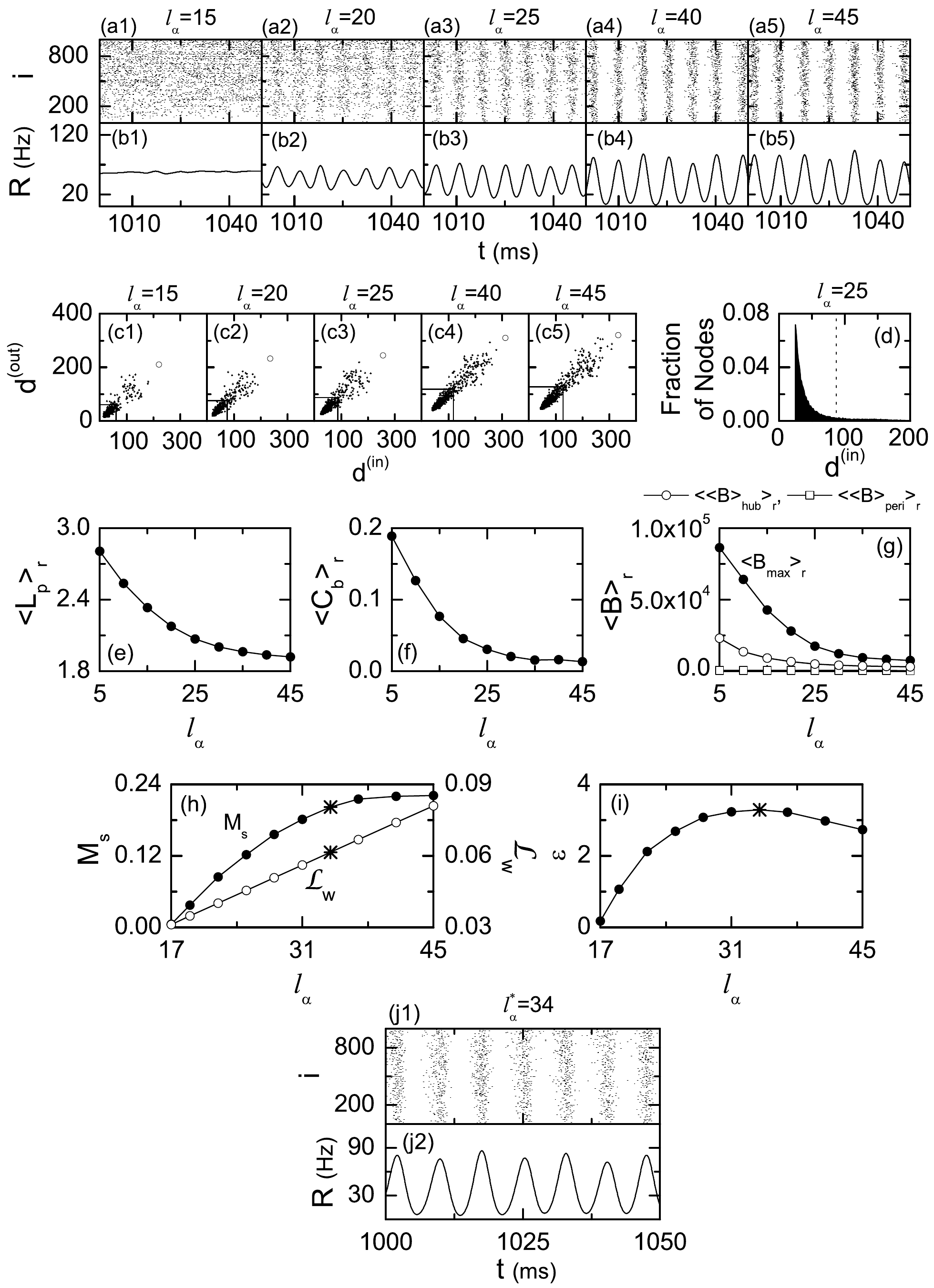}
\linespread{1.}
\caption{
Effect of symmetric attachment degree $l_\alpha$ on the sparse synchronization and economic SFN for $I_{DC}=1500$, $J=1500$, and $D=450$ in the directed SFN of $\alpha =1$ (i.e., $\beta=0$).
Raster plots of spikes in (a1)-(a5) and plots of the IPSR kernel estimate $R(t)$ in (b1)-(b5) for various values of symmetric attachment degree $l_{\alpha}$. The band width of the Gaussian kernel estimate for the IPSR $R(t)$ is 1 ms. Plots of the out-degree $d^{(out)}$ versus the in-degree $d^{(in)}$ for $l_\alpha=$ (c1) 15, (c2) 20, (c3) 25, (c4) 40, and (c5) 45. Peripheral groups are enclosed by rectangles, while hubs lie outside the rectangles. The head hub with the highest degree is represented by the open circle. (d) Histogram for fraction of nodes versus the in-degree $d^{(in)}$ for $l_\alpha=25$. This histogram is obtained through 30 realizations and the bin size for the histogram is 1. The vertical line represents a threshold for $d^{(in)}$ whose fraction of nodes is $0.002$ (i.e., $0.2\%$). Plots of (e) average path length $L_p$  and (f) betweenness centralization $C_b$ versus $l_{\alpha}$. (g) Plots of the maximum betweenness centrality $B_{max}$, the average betweenness centrality ${\langle B \rangle}_{hub}$ of secondary hubs, and the average betweenness centrality ${\langle B \rangle}_{peri}$ of peripheral nodes versus $l_{\alpha}$. Here, ${\langle \cdots \rangle}_r$ represents an average over 30 realizations. (h) Plots of statistical-mechanical spiking measure $M_s$ and normalized wiring length ${\cal{L}}_w$ versus $l_{\alpha}$. (i) Dynamical efficiency $\cal{E}$ versus $l_{\alpha}$. The values of $M_s$, ${\cal{L}}_w$, and $\cal{E}$ at an optimal value of $l_{\alpha}^*=34$ are denoted by the symbol ``*.'' Optimally fast synchronized rhythm for $l_{\alpha}=l_{\alpha}^*$: (j1) raster plot of neural spikes and (j2) plot of the IPSR kernel estimate $R(t)$.
}
\label{fig:ESFN}
\end{figure}

\newpage
\begin{figure}
\includegraphics[width=0.8\columnwidth]{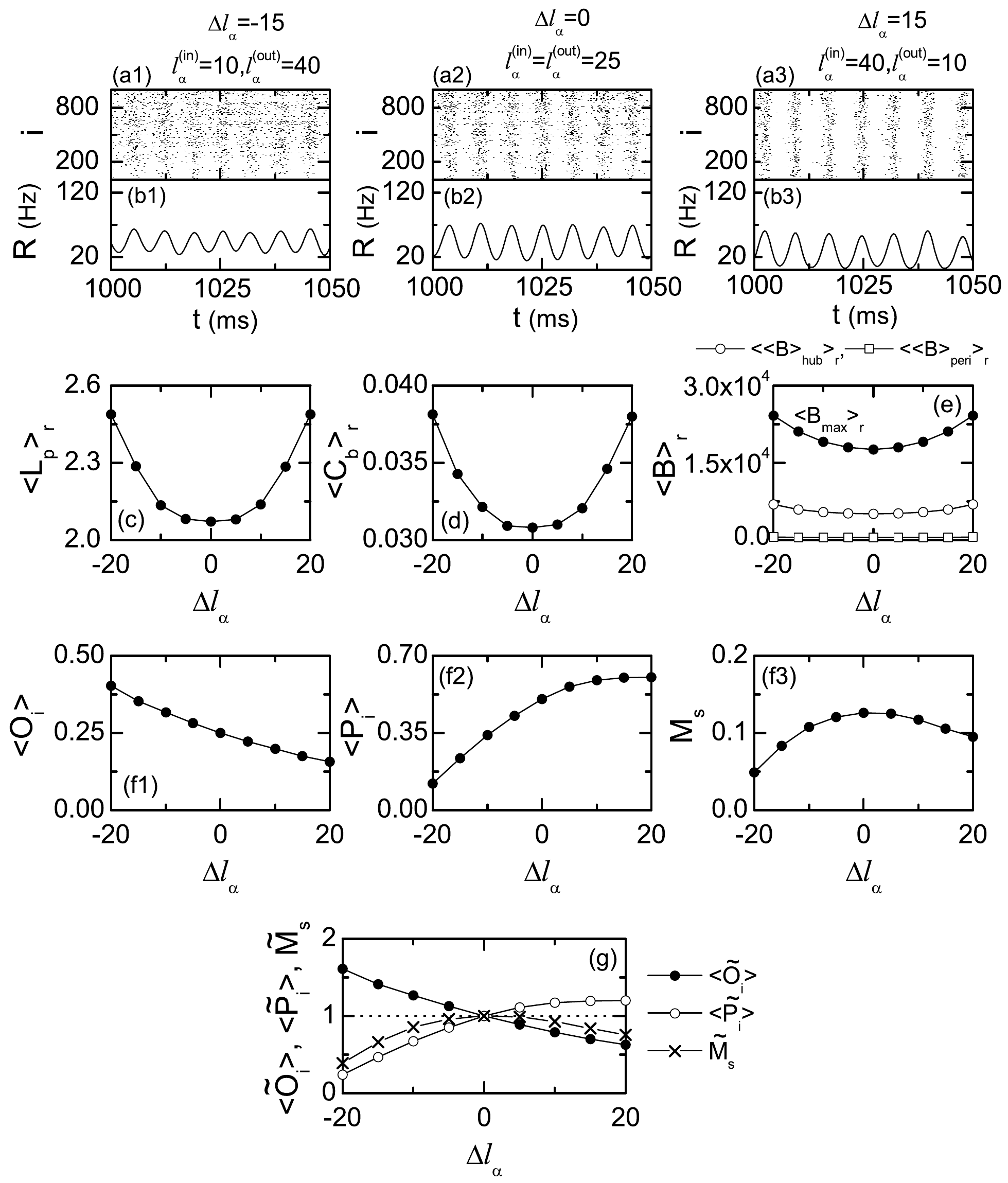}
\linespread{1.2}
\caption{Effect of asymmetric attachment on the sparse synchronization for $I_{DC}=1500$, $J=1500$, and $D=450$ in the directed SFN of $\alpha =1$ (i.e., $\beta=0)$, $l_{\alpha}^{(in)}=25 + \Delta l_{\alpha}$, and
$l_{\alpha}^{(out)}=25 - \Delta l_{\alpha}$ ($\Delta l_{\alpha}$: asymmetry parameter). Raster plots of spikes in (a1)-(a3) and plots of the IPSR kernel estimate $R(t)$ in (b1)-(b3) for various values of $\Delta l_{\alpha}$= -15, 0, and 15. The band width of the Gaussian kernel estimate for the IPSR $R(t)$ is 1 ms. Plots of (c) average path length $L_p$ and (d) betweenness centralization $C_b$  versus $\Delta l_{\alpha}$. (e) Plots of the maximum betweenness centrality $B_{max}$, the average betweenness centrality ${\langle B \rangle}_{hub}$ of secondary hubs, and the average betweenness centrality ${\langle B \rangle}_{peri}$ of peripheral nodes. Here, ${\langle \cdots \rangle}_r$ represents an average over 30 realizations. Plots of (f1) the average occupation degree $\langle O_i \rangle$, (f2) the average pacing degree $\langle P_i \rangle$, and (f3) the statistical-mechanical spiking measure $M_s$ versus $\Delta l_{\alpha}$. $\langle O_i \rangle$, $\langle P_i \rangle$, and $M_s$ are obtained by following $3 \times 10^3$ stripes in the raster plot of spikes. (g) Plots of the normalized average occupation degree $\langle {\widetilde{O}_i} \rangle$, the normalized average pacing degree $\langle {\widetilde{P}_i} \rangle$, and the normalized statistical-mechanical spiking measure $\widetilde{M}_s$. Normalizations of $\langle O_i \rangle$, $\langle P_i \rangle$,
and $M_s$ are done by dividing them with the values for the case of $\Delta l_\alpha=0$.
}
\label{fig:AA}
\end{figure}

\newpage
\begin{figure}
\includegraphics[width=\columnwidth]{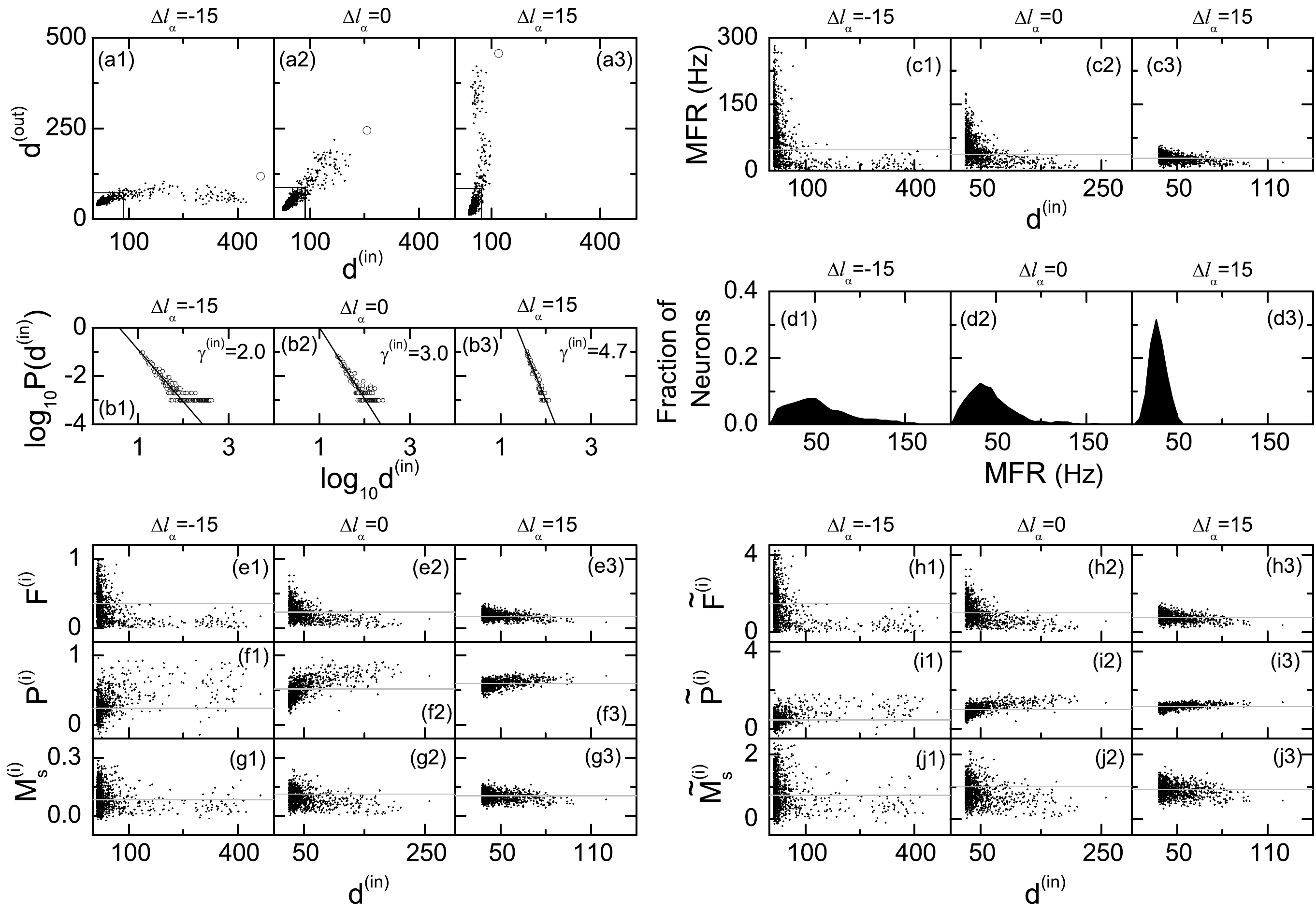}
\caption{Distinct differences in individual neuronal dynamics for the case of asymmetric attachment when $I_{DC}=1500$, $J=1500$, and $D=450$ in the directed SFN of $\alpha =1$ (i.e., $\beta=0$), $l_{\alpha}^{(in)}=25 + \Delta l_{\alpha}$, and $l_{\alpha}^{(out)}=25 - \Delta l_{\alpha}$ ($\Delta l_\alpha=$ -15, 0, and 15). (a1)-(a3) Plots of the out-degree $d^{(out)}$ versus the in-degree $d^{(in)}$ and (b1)-(b3) power-law in-degree distributions with different exponents for the cases of $\Delta l_{\alpha}=$ -15, 0, and 15. The fractions of nodes are $0.2 \%$ at the thresholds $d_{th}^{(in)}$ and $d_{th}^{(out)}$ of the in- and out-degrees which determine the rectangle enclosing the peripheral group. (c1)-(c3) Plots of MFRs versus the in-degree $d^{(in)}$ and (d1)-(d3) histograms for fraction of neurons versus MFR for the cases of $\Delta l_{\alpha}=$ -15, 0, and 15. Plots of (e1)-(e3) firing degree $F^{(i)}$, (f1)-(f3) pacing degree $P^{(i)}$, and (g1)-(g3) spiking measures $M_s^{(i)}$ for individual neurons versus the in-degree $d^{(in)}$ and plots of (h1)-(h3) normalized firing degree $\widetilde{F}^{(i)}$, (i1)-(i3) normalized pacing degree $\widetilde{P}^{(i)}$, and (j1)-(j3) normalized spiking measure $\widetilde{M}_s^{(i)}$ for the cases of $\Delta l_{\alpha}=$ -15, 0, and 15. Normalizations of $F^{(i)}$, $P^{(i)}$, and $M_s^{(i)}$ are done by dividing them with the values for the case of $\Delta l_\alpha=0$. Gray horizontal lines represent the ensemble-averaged values.
}
\label{fig:IDAA}
\end{figure}

\newpage
\begin{figure}
\includegraphics[width=0.7\columnwidth]{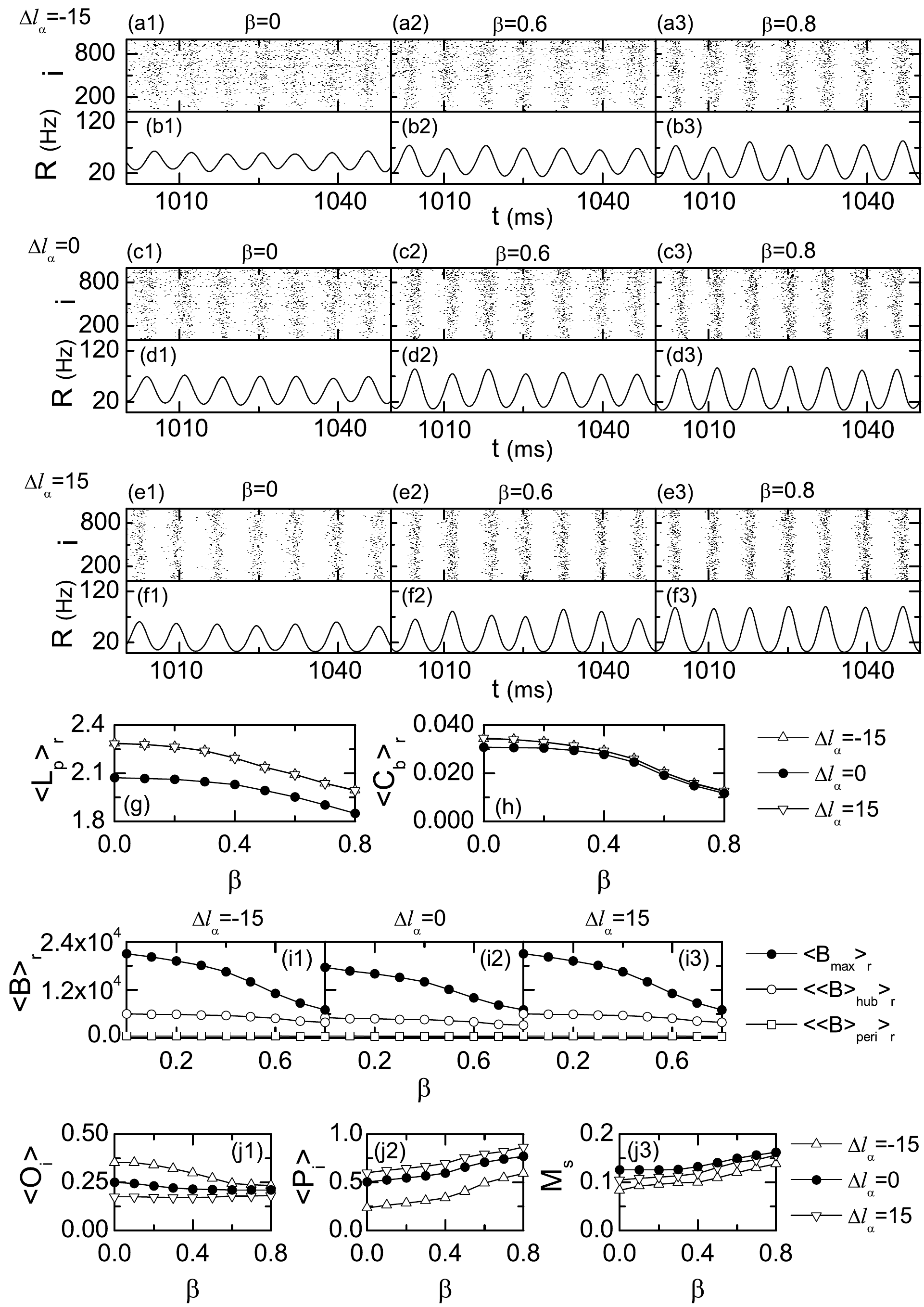}
\linespread{1.1}
\caption{Effect of $\beta$-process on sparse synchronization for $I_{DC}=1500$, $J=1500$, and $D=450$ in the directed SFN of $\beta =0$, 0.6, and 0.8
(i.e., $\alpha = 1 - \beta)$. For the $\alpha-$process, $l_{\alpha}^{(in)}=25 + \Delta l_{\alpha}$ and $l_{\alpha}^{(out)}=25 - \Delta l_{\alpha}$ ($\Delta l_\alpha=$ -15, 0, and 15), while for the $\beta-$process
$l_\beta^{(in)} = l_\beta^{(out)} \equiv l_\beta =5$. Effect of $\beta$-process for $\Delta l_{\alpha}=-15$: (a1)-(a3) Raster plots of spikes and (b1)-(b3) plots of the IPSR kernel estimate $R(t)$ for the cases of $\beta=0$, 0.6, and 0.8. Effect of $\beta$-process for $\Delta l_{\alpha}=0$: (c1)-(c3) Raster plots of spikes and (d1)-(d3) plots of the IPSR kernel estimate $R(t)$ for the cases of $\beta=0$, 0.6, and 0.8. Effect of $\beta$-process for $\Delta l_{\alpha}=15$: (e1)-(e3) Raster plots of spikes and (f1)-(f3) plots of the IPSR kernel estimate $R(t)$ for the cases of $\beta=0$, 0.6, and 0.8. Plots of (g) average path length $L_p$  and (h) betweenness centralization $C_b$ versus $\beta$ and (i1)-(i3) Plots of the maximum betweenness centrality $B_{max}$, the average betweenness centrality ${\langle B \rangle}_{hub}$ of secondary hubs, and the average betweenness centrality ${\langle B \rangle}_{peri}$ of peripheral nodes versus $\beta$ for the cases of $\Delta l_\alpha=$-15, 0. and 15. Here, ${\langle \cdots \rangle}_r$ represents an average over 30 realizations. Plots of (j1) average occupation degree $\langle O_i \rangle$, (j2) average pacing degree $\langle P_i \rangle$, and (j3) statistical-mechanical spiking measure $M_s$ versus $\beta$ for the three cases of $\Delta l_{\alpha}=$ -15, 0, and 15.
}
\label{fig:BP}
\end{figure}

\newpage
\begin{figure}
\includegraphics[width=\columnwidth]{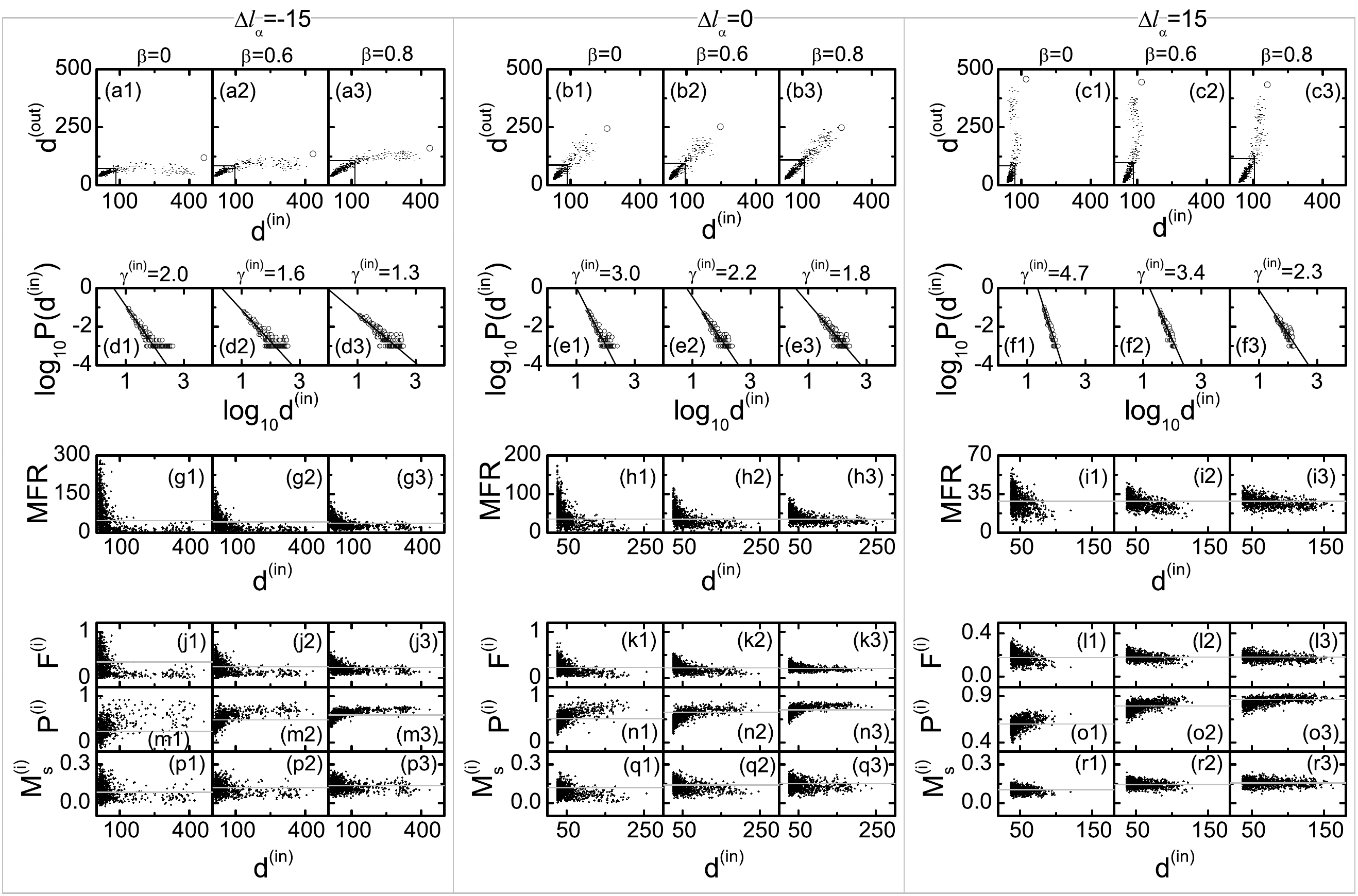}
\caption{Distinctly different individual dynamics in the $\beta$-process for $I_{DC}=1500$, $J=1500$, and $D=450$ in the directed SFN of $\beta =0$, 0.6, and 0.8 (i.e., $\alpha = 1 - \beta)$. For the $\alpha-$process, $l_{\alpha}^{(in)}=25 + \Delta l_{\alpha}$ and $l_{\alpha}^{(out)}=25 - \Delta l_{\alpha}$ ($\Delta l_\alpha=$ -15, 0, and 15), while for the $\beta-$process $l_\beta^{(in)} = l_\beta^{(out)} \equiv l_\beta =5$.
Plots of the out-degree $d^{(out)}$ versus the in-degree $d^{(in)}$ for $\Delta l_{\alpha}$= (a1)-(a3) -15, (b1)-(b3) 0, and (c1)-(c3) 15. The fractions of nodes are $0.2 \%$ at the thresholds $d_{th}^{(in)}$ and $d_{th}^{(out)}$ of the in- and out-degrees which determine the rectangle enclosing the peripheral group. Power-law in-degree distributions for $\Delta l_{\alpha}$= (d1)-(d3) -15, (e1)-(e3) 0, and (f1)-(f3) 15. Plots of MFRs versus the in-degree $d^{(in)}$ for $\Delta l_{\alpha}$= (g1)-(g3) -15, (h1)-(h3) 0, and (i1)-(i3) 15. Plots of firing degree $F^{(i)}$ for individual neurons versus the in-degree $d^{(in)}$ for $\Delta l_{\alpha}$= (j1)-(j3) -15, (k1)-(k3) 0, and (l1)-(l3) 15.
Plots of pacing degree $P^{(i)}$ for individual neurons versus the in-degree $d^{(in)}$ for $\Delta l_{\alpha}$= (m1)-(m3) -15, (n1)-(n3) 0, and (o1)-(o3) 15. Plots of spiking measure $M_s^{(i)}$ for individual neurons versus the in-degree $d^{(in)}$ for $\Delta l_{\alpha}$= (p1)-(p3) -15, (q1)-(q3) 0, and (r1)-(r3) 15.
}
\label{fig:IDBP}
\end{figure}

\end{document}